\documentclass[hyper,11pt,letterpaper]{JHEP3}
\usepackage[dvips]{epsfig}
\usepackage{amsmath,amssymb,epsf,amsfonts}
\usepackage[numbers,sort&compress]{natbib}
\usepackage{graphicx}
\usepackage{dcolumn}
\usepackage{bm}
\usepackage{amsthm}

\newcommand{\prn}[1]{\left ( #1 \right )}
\newcommand{\brk}[1]{\left [ #1 \right ]}

\title{Anomaly/Transport in an Ideal Weyl gas}
\author{\ R. Loganayagam$^a$\footnote{nayagam@physics.harvard.edu}\ ,\  Piotr Sur\'{o}wka$^b$\footnote{piotr.surowka@vub.ac.be}\\
\small{\emph{$^{a}$Junior Fellow, Harvard Society of Fellows,}} \\
\small{\emph{Harvard University, Cambridge, MA 02138 .}} \\ \\
\small{\emph{$^{b}$Theoretische Natuurkunde, Vrije Universiteit Brussel and}} \\
\small{\emph{The International Solvay Institutes,}} \\
\small{\emph{Pleinlaan 2, B-1050 Brussels, Belgium .}} 
}

\abstract{
We study some of the transport processes which are specific to an ideal gas of
relativistic Weyl fermions and relate the corresponding transport coefficients
to various anomaly coefficients of the system.  We propose that these
transport processes can be thought of as arising from the continuous injection of
chiral states and their subsequent adiabatic flow driven by vorticity. This in turn 
leads to an elegant expression relating the anomaly induced transport coefficients
to the anomaly polynomial of the Ideal Weyl gas.}

\preprint{}

\begin{document}

\section{Introduction}\label{sec:intro}

Anomalies are arguably among the most interesting phenomena to come out of
studies of quantum matter. Their importance lies in their robustness across
various length/energy scales - as one passes from one description of matter
into another, anomaly matching ala `t Hooft ensures that the underlying
anomalies of a theory survive in various disguises. 

While this statement is relatively better understood within the 
realm of effective theories (as exemplified by the phenomenology 
of  WZW term in particle physics and solid-state physics) , 
we have only a  limited understanding of the  role of 
anomalies in various finite  temperature/finite chemical
potential setups. Any progress in  the phenomenology of 
anomalies is welcome especially given the important role 
of quantum anomalies and their associated transport phenomena 
in fields ranging from solid-state physics to cosmology.

To be more precise, we are concerned with the following situation :
consider a quantum system with a continuous symmetry\footnote{
To simplify our discussion, we will assume this global symmetry is not spontaneously
broken - in other words, we are interested in transport processes
in normal fluid not superfluids. Though various ideas that we 
discuss in the article have their counterparts in superfluids,
we believe the phenomenology of anomaly-induced transport 
in superfluids is sufficiently different to merit a separate
discussion. There is by now a vast literature on such transport phenomena
which are beset with their own subtleties. Since it would be too
tedious/distracting to compare and contrast the effect of 
anomalies in the  two situations, we will choose entirely focus
on normal fluids  in what follows.  } which via
Noether theorem corresponds to conserved Noether currents. 
Consider gauging this symmetry by introducing a set of external
non-dynamical gauge fields. We will say the quantum system has an
anomaly if in the presence of such non-dynamical gauge fields 
the covariant Noether currents are no more conserved. One can now
consider instead turning on temperature/chemical potential for 
the Noether currents and ask what novel processes are characteristic of
a quantum system with underlying anomalies. 

By now, such transport processes have been studied from various points of 
view - they are known to be constrained by thermodynamics/adiabaticity in 
arbitrary dimensions \cite{Son:2009tf,Neiman:2010zi,Kharzeev:2011ds, Loganayagam:2011mu},
the corresponding transport coefficients can be derived via a 
Kubo-like formula in $4d$ \cite{Amado:2011zx}. They are  
an established feature in various holographic fluid phases in
CFTs dual to AdS$_3$ \cite{Kraus:2005zm} and AdS$_5$ \cite{Bhattacharyya:2007vs,
Banerjee:2008th,Erdmenger:2008rm,Torabian:2009qk, Landsteiner:2011iq}
where the CFT anomalies are in one to one correspondence with
the Chern-Simons terms in the AdS bulk. The effect of Chern-Simons term for
a U(1) gauge field in arbitrary  AdS$_{2n+1}$/CFT$_{2n}$ was 
worked out in \cite{Kharzeev:2011ds} - no analogous results are 
known for gravitational Chern-Simons terms in higher AdS spacetimes
\footnote{It would be interesting to construct and study rotating solutions
in AdS$_{2n+1}$ with pure/mixed gravitational Chern-Simons terms and link the
proposed modification of Wald entropy \cite{Tachikawa:2006sz,Bonora:2011gz} 
against the anomaly-induced entropy transport in the dual CFT.  }. 

In 2d field theories, the relation between U(1) anomaly and transport
is one the foundations of modern theories of Hall effects and there is 
an analogous relation between thermal transport and gravitational anomaly
(see for example \cite{2002NuPhB.636..568C}). There are by now various 
ways in which such transport processes in 4d free fermion theories have
been derived - some of them quite old \cite{Vilenkin:1978hb,Vilenkin:1979ui,
Vilenkin:1980fu,Vilenkin:1980zv,Vilenkin:1980ft,Vilenkin:1995um} and others more 
recent \cite{Fukushima:2008xe,Kharzeev:2009p,Landsteiner:2011cp}. Further, since $4d$ Weyl 
fermions are systems with Berry phases, these transport processes have close
links to the general theory of Berry phases and transport \cite{2004PhRvL..93t6602H,
2010RvMP...82.1959X} and in particular transport in Weyl semi-metals \cite{2011PhRvB..83t5101W,
2011PhRvB..84g5129Y}. The transport processes linked to anomalies have 
also made their appearance in discussions about 
classification of topological insulators \cite{2010arXiv1010.0936R,Hughes:2011hv}.

At present, the most general set of results were derived via thermodynamic
arguments employing adiabaticity\cite{Kharzeev:2011ds,Loganayagam:2011mu}.
These results can be summarized as follows \footnote{We will use the notations of 
\cite{Loganayagam:2011mu} in the following. See section \S\ref{sec:setup}
and the appendix \S\ref{app:notation} for a discussion of the basic setup
and notations.} : In a fluid the energy, charge and 
entropy transport are given by 
\begin{equation}
\begin{split}
T^{\mu\nu} &\equiv \varepsilon u^\mu u^\nu + p P^{\mu\nu} + q^\mu_{anom}u^\nu + u^\mu q^\nu_{anom} + T^{\mu\nu}_{diss}\\
J^{\mu} &\equiv n u^\mu + J^{\mu}_{anom}+J^{\mu}_{diss} \\
J^\mu_S &\equiv s u^\mu + J^\mu_{S,anom}+J^\mu_{S,diss}\\
\end{split}
\end{equation}
where $u^\mu$ is the velocity of the fluid under consideration which obeys $u^\mu u_\mu =-1$ when
contracted using the spacetime metric $g_{\mu\nu}$. Further, $P^{\mu\nu}\equiv g^{\mu\nu}+u^\mu u^\nu$ , 
pressure of the fluid is $p$ and $\{\varepsilon,n,s\}$ are the 
energy,charge and the entropy densities respectively. We have denoted by $\{q^\mu_{anom},J^{\mu}_{anom},
J^\mu_{S,anom}\}$ the anomalous heat/charge/entropy currents and by $\{T^{\mu\nu}_{diss},J^{\mu}_{diss},
J^\mu_{S,diss}\}$ the dissipative currents.

We are primarily interested in the anomalous currents in 
what follows. It is convenient to work with forms - let 
$\{\bar{q}_{anom},\bar{J}_{anom},\bar{J}_{S,anom}\}$ be the 
Hodge duals of the corresponding currents\footnote{Throughout this
article, we will use overbars to denote Hodge duals.}. Using adiabaticity the following statements can 
be made in flat spacetime\cite{Loganayagam:2011mu} 
\begin{enumerate}
\item All these currents are derivable from a single Gibbs current
$\bar{\mathcal{G}}_{anom}$ which describes the transport of Gibbs free energy 
$G\equiv E-TS-\mu Q$ in the fluid. We have the following thermodynamic relations
\begin{equation}
\begin{split}
\bar{J}_{anom} &=-\frac{\partial \bar{\mathcal{G}}_{anom}}{\partial \mu}\\
\bar{J}_{S,anom}  &=-\frac{\partial \bar{\mathcal{G}}_{anom}}{\partial T} \\
\bar{q}_{anom} &= \bar{\mathcal{G}}_{anom}+T\bar{J}_{S,anom}+\mu \bar{J}_{anom}\\
\end{split}
\end{equation}
\item  $\bar{\mathcal{G}}_{anom}$ is determined in terms of the fluid 
vorticity 2-form $\omega$ , the rest-frame magnetic field 2-form $B$
and a $(n+1)^{th}$degree polynomial $\mathfrak{F}^\omega_{anom}[T,\mu]$
in temperature $T$ and chemical potential $\mu$
(where $d=2n$ is the number of spacetime dimensions). The explicit expression for 
 $\bar{\mathcal{G}}_{anom}$  is given by\footnote{This is just a convenient rephrasing
of the formulae/results presented in the Appendix \S A of \cite{Loganayagam:2011mu}.
In particular, we have the following relation relating the functions appearing there
to the functions appearing here
\[\mathfrak{F}^\omega_{anom}[T(2\omega),B+\mu(2\omega)] = \mathfrak{f}[B+\mu(2\omega)]+\frac{1}{2}T^2(2\omega)^2\mathfrak{g}[B+\mu(2\omega),T\omega] \]
 }

\begin{equation}\label{eqn:Fw}
\begin{split}
\bar{\mathcal{G}}_{anom}
&=\frac{1}{(2\omega)^2}\Bigl\{\mathfrak{F}^\omega_{anom}[T(2\omega),B+\mu(2\omega)]-\Bigl[\mathfrak{F}^\omega_{anom}[T(2\omega),B+\mu(2\omega)]\Bigr]_{\omega=0}
 \Bigr.\\
&\qquad\qquad\qquad\Bigl.-\quad\omega\Bigl[\frac{\delta}{\delta\omega}\mathfrak{F}^\omega_{anom}[T(2\omega),B+\mu(2\omega)]\Bigr]_{\omega=0} \Bigr\}\wedge u\\ 
\end{split}
\end{equation}
\item The  polynomial $\mathfrak{F}^\omega_{anom}[T,\mu]$ obeys two constraints - first, it has no term linear in $T$. Second,
its value at zero temperature is completely determined by the $U(1)$ anomaly in the system. 
\item To these statements, we can add the following statement which does not follow from thermodynamic
arguments in  \cite{Loganayagam:2011mu} but nevertheless seems to be true across various systems - the $T^2$ coefficient in $2d$ and $4d$ seem to be related to gravitational anomalies (see for example \cite{Landsteiner:2011cp,Landsteiner:2011iq} for
results in $4d$.
\end{enumerate}

This summary encompasses everything that is known till now about these transport processes just from 
flat spacetime thermodynamics alone. But, this is unsatisfactory for various reasons - first of
all, as we had mentioned various gravitational anomalies of the system show up even in the 
flat spacetime transport and this is mysterious even from the point of view of flat spacetime
thermodynamics. Second, we do not know how these relations to gravitational anomalies generalize
to higher dimensions. Further, a more microscopic understanding of these transport processes
would clearly be useful for various reasons - for example, we would like
to study these  transport processes away from equilibrium/in the presence of disorder/
in the lattice analogues of continuum Weyl fermions. Our aim in this article is to begin 
addressing these questions in the simplest system exhibiting such transport - free fermion theory in 
even spacetime dimensions $d=(2n-1)+1$ with a collection of fermions with different 
chiralities $\chi_{_{d=2n}}$ and charges $q$ under some specific $U(1)$ global symmetry
of the free theory.

After establishing the basic setup/notations in the section
\S\ref{sec:setup}, we proceed to study in detail the simplest case of $1+1$d 
chiral fermions in flat spacetime in section \S\ref{sec:2d}. We then translate the thermodynamic
arguments of adiabaticity in flat spacetime into a more microscopic set of equations which one can
take as an alternate starting point for the $2d$ analysis and which easily 
generalizes to higher dimensions. We propose the following intuitive picture for
how the transport phenomena linked to anomaly arise : first of all, we propose that
these transport processes can be thought of as arising from a certain \emph{chiral}
density of states and their \emph{spectral flow}. In a continuum description, we capture this
by a \emph{chiral spectral current} $\mathcal{J}^\mu_q(x,E_p)$ whose 
time component is the chiral density of 1-particle states with charge $q$
and the rest-frame energy $E_p$ and whose spatial components tell us about the flow of 
such states as the fluid flows. In section \S\ref{sec:adiab}, we argue that
adiabaticity can be simply seen as a conservation type equation for this 
chiral spectral current.

This is of course a very well-known aspect of anomalies whereby 
the basic process driving the anomaly is the
continuous injection of chiral zero modes by the external magnetic field 
(which we will assume to be small and slowly varying) into the system.
We make this precise in  section \S\ref{sec:adiabAnom} by simply 
matching the microscopic discussion of adiabaticity to the thermodynamic
discussion of adiabaticity. This matching gives a boundary
condition for the conservation equation for the chiral spectral current by giving 
the rate at which chiral modes are injected into the fluid.
\begin{equation}
\begin{split}
\bar{\mathcal{J}}_q|_{E_p=0} &= \frac{\chi_{_{d=2n}}}{2\pi}    \prn{\frac{qB}{2\pi}}^{n-1}\wedge \frac{u}{(n-1)!}\\
\end{split}
\end{equation}
These modes, once injected by the magnetic field are then convected along with the 
fluid flow and the adiabaticity is just the statement that there is no more
creation/annihilation of states at finite energies.

Armed with this intuitive picture, we proceed in section \S\ref{sec:solvAdiab} to solve the 
conservation equation in flat spacetime. This results in a simple expression for the chiral spectral 
current
\begin{equation}
\begin{split}
\bar{\mathcal{J}}_q &= \frac{\chi_{_{d=2n}}}{2\pi}    \prn{\frac{qB+2\omega E_p}{2\pi}}^{n-1}\wedge \frac{u}{(n-1)!}\\
\end{split}
\end{equation}
where $\chi_{_{d=2n}}$ is the chirality of the 1-particle state. This solution tells us
how the states of different energies and charges flow hence
solving once for all the spectral flow problem in an arbitrary fluid flow (in flat spacetime).
We now notice a remarkable result - if we take the  external field strengths
to zero $B\to 0$, there is no more an injection of new zero-energy states into the
fluid, but the chiral spectral flow is still non-zero even if the anomaly is turned off !
In this case, the vorticity is sufficient to drive the chiral spectral 
current and this is the basic reason why rotational response encodes information
about the anomalies in the system - both gravitational and non-gravitational.

In the next section \S\ref{sec:FwSoln}, we add up the Gibbs-free energy contribution of
each 1-particle state to get the anomaly-induced free-energy current 
$\bar{\mathcal{G}}_{anom}$. We find that $\bar{\mathcal{G}}_{anom}$ is of the form
given in eqn.\eqref{eqn:Fw} which was derived in \cite{Loganayagam:2011mu} by 
thermodynamic considerations. While this is not surprising, we find the 
the polynomial $\mathfrak{F}^\omega_{anom}$ is derived by a very simple
formula from the anomaly polynomial of the system.  We get
\begin{equation}
\begin{split}
{\mathfrak{F}}^\omega_{anom}&=\mathcal{P}_{anom}\prn{F\mapsto \mu ,\ p_{_1}(\mathfrak{R}) \mapsto -T^2,\ p_{_{k>1}}(\mathfrak{R}) \mapsto  0 }
\end{split}
\end{equation}
where $\mathcal{P}_{anom}$ is  the anomaly polynomial of the system written in terms of the 
$U(1)$ field strength $F$ and the $k^{th}$ Pontryagin forms\footnote{See appendices
\S\ref{app:anom} for definitions of various quantities related
to anomaly polynomials.} of the spacetime curvature $p_{_k}(\mathfrak{R})$. The above formula
gives a simple replacement rule by which one can go from the anomaly polynomial to the
polynomial ${\mathfrak{F}}^\omega_{anom}$. 

Note that this a generalization of the observation made in \cite{Landsteiner:2011cp}  that  $T^2$ 
coefficient in $4d$ free theories seem to be related to gravitational anomalies.
Since the $4d$ relation continues to hold in strongly coupled holographic phases
too \cite{Landsteiner:2011iq}, it is tempting to conjecture that the above replacement 
rule would continue to hold even beyond free theories. We emphasize that this is quite
a non-trivial conjecture and what we have is a preliminary evidence that it might be true.
We discuss this along with other further directions in our discussion section \S\ref{sec:disc}.
We collect various useful results in our appendices. In the next section, we begin
by explaining our basic setup and defining our notation - most of it being quite standard and 
elementary, the reader should feel free to skim through the section just noting various remarks
on notation.

\section{The Basic setup}\label{sec:setup}
The main system we will be concerned about throughout this article is a system of free
relativistic fermions at finite temperature and chemical potential  in a flat spacetime 
with the spacetime dimension $d=2n=(2n-1)+1$ being even. Every particle state 
occurs along with its anti-particle state and we will call such a
particle/anti-particle pair as a \emph{species}. Hence, we will denote by $\sum_{species}$
the summation where each particle/anti-particle pair contributes a single term to the sum.
This should be distinguished from $\sum_F$ which denotes the summation where each particle
(or anti-particle) state contributes separately to the sum.

Among the large symmetry that the free theory enjoys, we will choose a specific
$U(1)$ symmetry for which we will turn on the chemical potential
$\mu$ and put the system at a finite temperature $T$ (we will use the letter 
$\beta\equiv 1/T$ to denote the inverse temperature) . The thermodynamic $T$ and $\mu$
are of course defined on a local rest frame defined via a unit time-like vector 
$u^\mu$. The thermodynamic potential appropriate to such a grand-canonical 
description is the Gibbs free-energy density $\mathcal{G}\equiv \varepsilon-Ts-\mu n$
where $\varepsilon,s,n$ are the  rest-frame energy density, entropy density and the charge 
density respectively. The first law takes the form $d\mathcal{G}=-sdT- nd\mu$. 

For a free theory, the Gibbs free-energy is obtained by simply multiparticling 
contribution from the 1-particle sector. We will denote the Gibbs free-energy
contribution of a  fermionic state with a charge $q$ and rest frame energy $E_p$
by $g_q$. Hence
\[ g_q \equiv -\frac{1}{\beta} \ln\brk{1+ e^{-\beta\prn{E_p-q\mu}}} \]  
The occupation of a given 1-particle fermionic state is given by the Fermi-Dirac
distribution denoted by $f_q$
\[ f_q \equiv \frac{1}{e^{\beta\prn{E_p-q\mu}}+1} \]
The contribution of a 1-particle state to the entropy is given by the \emph{negative}
of the Boltzmann's $\mathcal{H}$-function which for fermionic states takes the form 
\[ \mathcal{H}_q \equiv f_q\ln f_q + (1-f_q)\ln\prn{1-f_q} \]

Using the standard terminology , we will call a state with slowly varying 
$T,\mu,u^\mu$ with the local density matrix being close to the thermal 
density matrix  as an \emph{Ideal gas}. Since the constituents are Weyl
fermions, we will call this an Ideal Weyl gas. In this article, we will
be concerned about a specific subset of transport processes which are linked
to various anomalies in the ideal Weyl gas. Before proceeding let us dispose
of a specific technicality : as is well known, in the strictly non-interacting
limit, all the dissipation length/time scales diverge which in turn means that
dissipative transport coefficients like viscosity also diverge in this limit.
While this is true, this shall not worry us too much since the effects that we
are looking for are non-dissipative and are not plagued by such `free-theory'
infinities. While we expect addition of interactions/dissipation would not
modify our results (this expectation is partially justified by various existent
calculations in strongly coupled holographic phases), it would be nice to
explicitly prove this statement.

Having addressed that technicality, let us continue : we want to study such an
ideal Weyl gas in presence of non-dynamical background electric and magnetic fields.
Let $F$ be the field-strength 2-form, we define the rest frame electric 1-form 
via $E_{\mu}\equiv F_{\mu\nu} u^\nu $. We can then do an electric-magnetic decomposition
\begin{equation} F_{\mu\nu} -\left[ u_{\mu}E_{\nu}-E_{\mu}u_{\nu}\right] \equiv B_{\mu\nu} \end{equation}
or in the language of forms
\begin{equation} F = B + u\wedge E  \end{equation}
where $B$ is the rest-frame magnetic 2-form completely transverse to $u^\mu$, i.e., $B_{\mu\nu}u^\nu=0$.
We will also use the standard decomposition of the velocity gradients
\begin{equation} D_\mu u_\nu = \sigma_{\mu\nu}+\omega_{\mu\nu}-u_{\mu} a_\nu +\frac{\theta}{d-1}P_{\mu\nu} \end{equation}
in terms of the shear strain rate $\sigma_{\mu\nu}$, the vorticity $\omega_{\mu\nu}$,
the acceleration $a_\mu$ and the expansion rate $\theta$ of the fluid. Further 
$P_{\mu\nu}\equiv g_{\mu\nu}u_\mu u_\nu $ as before. This in 
particular means the exterior derivative of the velocity 1-form has
the decomposition
\begin{equation} Du = 2\omega -u\wedge a \end{equation}
where $\omega$ is the vorticity 2-form.  

Before we enter the main argument of the paper, we will make some comments 
regarding our conventions for chirality in higher dimensions. Consider
the 1-particle states which are given by the solutions of the 
Weyl equation of appropriate chirality (which is just the massless Dirac equation
with the opposite chirality projected out). We will define our conventions for
chirality now by essentially equating it to the helicity. To do this, let us divide the $2n-1$ spatial directions
into a direction $x_1$ and $n-1$ planes  where the $k^{th}$ plane is the
$(x_{i_{2k}},x_{i_{2k+1}})$ plane where $k=1,\ldots,n-1$. 

Consider first the positive frequency solutions of the Weyl equation. Further 
we will consider only the solutions with only $p^0>0,p^1\neq 0$ -
all other components of momentum being zero and the spin along 
$k^{th}$ plane being $S_{i_{2k} i_{2k+1}}\equiv\frac{1}{2}\sigma_k$  with
$\sigma_k=\pm 1$. Any other solution can be obtained by rotating this solution
or by linear combinations thereof.  For these solutions, we will assign the 
chirality via their helicity 
\[ \chi_{_{d=2n}}\equiv \text{sign}(p_1) \prod_{k=1}^{n-1} \sigma_k  \]
For a Weyl equation of a particular chirality $\chi_{_{d=2n}}$ only the $\sigma_k$
obeying the above equation are allowed. When $n=1$ (when $d=1+1d$) this gives a 
single state with the sign of $p_1$ fixed. For $n\geq 2$, we get $2^{n-2}$ states 
with $p_1$ being arbitrary. 

We now turn to the negative frequency solutions  which are complex conjugates of positive frequency anti-particle solutions. 
The complex conjugate of a Weyl spinor of chirality $\chi_{_{d=2n}}$ is 
another Weyl spinor of chirality $(-1)^{n-1}\chi_{_{d=2n}}$. Hence, for 
corresponding to every particle state above we get an anti-particle  state
with  chirality $(-1)^{n-1}\chi_{_{d=2n}}$. We will call the Weyl fermions
with positive chirality as left fermions and those with negative chirality 
as right fermions. This summarizes the basic definitions needed
for the rest of the paper. The reader can find a table of notation in the appendix
\S\ref{app:notation} for ready reference.

\section{Anomaly and transport in 2d Weyl gas}\label{sec:2d}
Before trying to tackle the case of a Weyl gas in higher dimensions, it is 
instructive to work out the simplest case of $1+1$ dimensions. This is a very
well-studied system and in some sense we will not have anything new\footnote{See
for example \cite{Dubovsky:2011sk} for a different take on anomaly/transport in 2d
fluids.}  to add except
for a way of looking at the standard results which will prepare us for the more
subtle effects in higher dimensions. With this objective in mind, we will 
focus on this simple case in some detail.

We will begin by considering a single species of a free left Weyl fermion in $1+1$ dimensions with 
charge $q$. We will assign this fermion a chirality $\chi_{_{d=2}}=+1$.  The
anti-particle of this fermion is again a left Weyl fermion with charge $-q$.
This follows from the general rule that in $(2n-1)+1$d the 
chirality of the anti-particle is $(-1)^{n+1}$ times the 
chirality of the particle.

We are interested in a gas of such Weyl fermions at a finite temperature $T$
and chemical potential $\mu$. These quantities are of course defined in 
a center of mass frame of the ideal Weyl gas - let this frame be defined by
a $1+1$-d unit time-like vector $u^\mu$ which we will take it to be constant.
It is this $u^\mu$ which in the hydrodynamic description will describe the 
fluid velocity.

The question we want to address is this - what is the hydrodynamic 
description of such an ideal gas ? We will first
give a naive answer to this question which will later correct.

The conventional intuition is that this system behaves like an 
ideal fluid with the following naive constitutive relations for 
energy/charge/entropy currents
\begin{equation} 
\begin{split}
T^{\mu\nu}_{\text{naive}} &= \varepsilon u^\mu u^\nu + p \prn{g^{\mu\nu}+u^\mu u^\nu}\\
J^\mu_{\text{naive}} &= n u^\mu \\
J^\mu_{S,\text{naive}} &= s u^\mu \\
\end{split}
\end{equation}
where  the energy density $\varepsilon$, pressure $p$ , charge density $n$ and 
entropy density $s$ can be calculated from the usual statistical mechanics of an
ideal fermion gas. We will calculate this in a moment, but before that we will
argue that the above form is definitely incomplete! 

The reason is simple - a theory of a free Weyl fermion is a holomorphic 2d CFT and hence only the 
holomorphic components of the  currents can be non-zero. The above relations
are in clear contradiction with holomorphy - for one, the charge/entropy currents 
are time-like rather than null as would be predicted by holomorphy. So we are 
led to the surprising statement that conventional semi-classical intuition about
the ideal gas is in direct contradiction with holomorphy in this simple system.
Having concluded thus, let us actually calculate carefully what the actual
constitutive relations should be.

A left Weyl fermion field in $2d$ is just a single component complex 
field $\psi$ which obeys the Weyl equation (or the massless 
Dirac equation)
\[ \brk{\partial_t+\partial_x} \psi=0 \]
This follows from the particular choice for  the Gamma 
matrices $\{\Gamma^t,\Gamma^x\}=\{-i\sigma_y,\sigma_x\}$.
We can repeat the same exercise for the right Weyl fermion with $\chi_{_{d=2}}=-1$ where we 
just flip the sign of the  $\partial_x$ term.  In the following we will write  
our formulae in such a way that the expressions for the right
Weyl fermion can be obtained by putting $\chi_{_{d=2}}=-1$. Hence
we write the Weyl equation as
\[ \brk{\partial_t+\chi_{_{d=2}}\partial_x} \psi=0 \]
This equation is easily solved - solutions are just plane-waves that travel from left to right in the space
\[ \psi = \int_0^\infty \frac{dE_p}{2\pi}\frac{1}{\sqrt{2E_p}}  \brk{a_p e^{ip.x} + b_p^\dag e^{-ip.x}}_{p^\mu =E_p\{1,\chi_{_{d=2}}\}} \]
where $a_p^\dag$ and $b_p^\dag$ are the creation operators for the particle and the anti-particle
respectively and $E_p$ is the energy in some arbitrary frame. 

Let us work in the rest frame define by $u^\mu$ from now on - so we take $u^\mu=\{1,0\}$. 
Let  $\epsilon^{\mu\nu}$ be the completely antisymmetric tensor in 2d with $\epsilon^{tx}=1$ which
implies $\epsilon^{\mu\nu}u_\nu= \{0,1\}$. The Weyl equation is
\[ \brk{u^\mu+\chi_{_{d=2}}\epsilon^{\mu\nu}u_\nu}\partial_\mu \psi=0 \]
and the plane-wave solutions above are 
\[ \psi = \int_0^\infty \frac{dE_p}{2\pi}\frac{1}{\sqrt{2E_p}} \brk{a_p e^{ip.x} + b_p^\dag e^{-ip.x}}_{p^\mu =E_p\brk{u^\mu+\chi_{_{d=2}}\epsilon^{\mu\nu}u_\nu}} \]
We want to populate the states of these fermions/anti-fermions in this frame according to 
the Fermi-Dirac distribution and calculate the conserved currents in the thermal ensemble. 
This gives
\begin{equation} 
\begin{split}
T^{\mu\nu} &= \sum_{species}\int_0^\infty \frac{dE_p}{2\pi} \prn{f_q + f_{-q}} E_p \brk{u^\mu+\chi_{_{d=2}}\epsilon^{\mu\alpha}u_\alpha}\  \brk{u^\nu+\chi_{_{d=2}}\epsilon^{\nu\lambda}u_\lambda} \\
&= \varepsilon u^\mu u^\nu + p \prn{g^{\mu\nu}+u^\mu u^\nu} + q^\mu_{anom} u^\nu + q^\nu_{anom} u^\mu\\
J^\mu &= \sum_{species}\int_0^\infty \frac{dE_p}{2\pi} \prn{qf_q -qf_{-q}}\brk{u^\mu+\chi_{_{d=2}} \epsilon^{\mu\alpha}u_\alpha}  \\
&= n u^\mu + J^\mu_{anom} \\
J^\mu_S &= -\sum_{species}\int_0^\infty \frac{dE_p}{2\pi} \prn{\mathcal{H}_q +\mathcal{H}_{-q}}\brk{u^\mu+\chi_{_{d=2}}\epsilon^{\mu\alpha}u_\alpha}  \\
&= s u^\mu + J^\mu_{S,anom} \\
\end{split}
\end{equation}
where we have used the relation $\chi_{_{d=2}}\epsilon^{\mu\alpha}u_\alpha\  \chi_{_{d=2}} \epsilon^{\nu\lambda}u_\lambda = g^{\mu\nu}+u^\mu u^\nu$.
We have collected together the deviations from the conventional hydrodynamic expectation under the objects \
with the subscript $anom$. We get  the conventional expressions which could have been naively guessed
\begin{equation} 
\begin{split}
\varepsilon &= p= \sum_{species}\int_0^\infty \frac{dE_p}{2\pi} \prn{f_q + f_{-q}} E_p = \sum_F \int_0^\infty \frac{dE_p}{2\pi} f_q  E_p \\
n &= \sum_{species}\int_0^\infty \frac{dE_p}{2\pi} \prn{qf_q -qf_{-q}}= \sum_F \int_0^\infty \frac{dE_p}{2\pi} qf_q  E_p \\
s &= -\sum_{species}\int_0^\infty \frac{dE_p}{2\pi} \prn{\mathcal{H}_q +\mathcal{H}_{-q}}= -\sum_F \int_0^\infty \frac{dE_p}{2\pi} \mathcal{H}_q\\
\end{split}
\end{equation}
where the sum is over every fermionic species with particles and antiparticles counted separately. 
The anomalous corrections are given by 
\begin{equation} 
\begin{split}
q^\mu_{anom} &= \sum_F \chi_{_{d=2}}\epsilon^{\mu\alpha}u_\alpha  \int_0^\infty \frac{dE_p}{2\pi} f_q E_p\\
J^\mu_{anom}  &= \sum_F \chi_{_{d=2}}\epsilon^{\mu\alpha}u_\alpha \int_0^\infty \frac{dE_p}{2\pi} qf_q  \\
J^\mu_{S,anom}  &= -\sum_F \chi_{_{d=2}}\epsilon^{\mu\alpha}u_\alpha \int_0^\infty \frac{dE_p}{2\pi} \mathcal{H}_q   \\
\end{split}
\end{equation}

At this point, the curious reader might wonder what happens to these relations
in the case of a Dirac fermion. The massless Dirac fermion is just a left 
Weyl fermion of charge $q$ along with a right Weyl fermion of charge $q$. 
In this case, it is evident from the expressions above that
the conventional terms add up and the anomalous terms cancel out. Hence the naive 
guess turns out to be correct for a massless Dirac fermion. It is not very 
difficult to convince oneself by  explicit computation that the 
naive guess works even for the massive Dirac fermion. This then is the first
lesson from this exercise : \emph{ there are transport processes in hydrodynamics
to which only chiral species contribute }.

Since we will be studying these anomalous contributions in more detail- let us 
simplify them by writing the currents above as 1-forms $\{q_{anom},J_{anom},J_{S,anom}\}$.
Further take a Hodge-dual on both sides (which we will denote by an overbar)
to remove the $\epsilon^{\mu\nu}$ to get
\begin{equation} 
\begin{split}
\bar{q}_{anom} &= \sum_F   \int_0^\infty \frac{dE_p}{2\pi} f_q E_p\ \chi_{_{d=2}}u\\
\bar{J}_{anom}  &= \sum_F  \int_0^\infty \frac{dE_p}{2\pi} qf_q \ \chi_{_{d=2}}u  \\
\bar{J}_{S,anom}  &= -\sum_F \int_0^\infty \frac{dE_p}{2\pi} \mathcal{H}_q\ \chi_{_{d=2}}u   \\
\end{split}
\end{equation}
These are easily calculated from the Gibbs free-energy current 
\begin{equation}
\begin{split}
\bar{\mathcal{G}}_{anom} &= \sum_F  \int_0^\infty \frac{dE_p}{2\pi} g_q\ \chi_{_{d=2}}u  \\
\bar{J}_{anom} &=-\frac{\partial \bar{\mathcal{G}}_{anom}}{\partial \mu}\quad ,\quad
\bar{J}_{S,anom}  =-\frac{\partial \bar{\mathcal{G}}_{anom}}{\partial T}\\
& \quad\text{and} \quad \bar{q}_{anom}= \bar{\mathcal{G}}_{anom}+T\bar{J}_{S,anom}+\mu \bar{J}_{anom}\\
\end{split}
\end{equation}
This means we basically have to evaluate only one thermal integral. This can be
done by pairing up the particle and anti-particle contribution and using the identity \footnote{See appendix
\S\ref{app:fermifn} for a derivation of this/related identities. }
\[   \int_0^\infty \frac{dE_p}{2\pi} \prn{g_q+g_{-q}} =  -2\pi\brk{\frac{1}{2!}\prn{\frac{q\mu}{2\pi}}^2+\frac{T^2}{4!}} \]
This gives
\begin{equation}
\begin{split}
\bar{\mathcal{G}}_{anom} &= - 2\pi\brk{\frac{\mu^2}{2!(2\pi)^2}\prn{\sum_{species} \chi_{_{d=2}} q^2}+\frac{T^2}{4!}\prn{\sum_{species} \chi_{_{d=2}}}} u\\
\end{split}
\end{equation}
where the sum is performed over the fermionic species with a particle-antiparticle pair contributing to
a single term in the sum. Now we note a crucial fact : the anomalous contribution is completely 
proportional to the $U(1)$ anomaly coefficient $\sum_{species} \chi_{_{d=2}} q^2$ and the Lorentz anomaly
coefficient  $\sum_{species} \chi_{_{d=2}}$. So, we come to the second lesson of this 
exercise : \emph{The anomalous transport is completely determined by the anomalies in the
system}. Hence, we will call such a transport as \textit{anomaly-induced}.

Let us make this more precise - the anomaly coefficients of a system are neatly summarized
by a polynomial in gauge field strength $F$ and spacetime curvature $\mathfrak{R}$. For a
collection of $2d$ Weyl fermions the anomaly polynomial is given by (see the
 appendix\S\ref{app:anom} for a review of anomaly polynomials for fermions) 
\begin{equation}
\begin{split}
\mathcal{P}_{anom}(F,\mathfrak{R})&\equiv  -2\pi\brk{\frac{F^2}{2!(2\pi)^2}\prn{\sum_{species} \chi_{_{d=2}} q^2}-\frac{p_{_1}(\mathfrak{R})}{4!}\prn{\sum_{species} \chi_{_{d=2}}}}_{2d}
\\
\end{split}
\end{equation}
where $p_{_1}(\mathfrak{R})$ is the first Pontryagin class of spacetime curvature defined as
\[p_{_1}(\mathfrak{R}) \equiv  -\frac{1}{2(2\pi)^2} \mathfrak{R}_{a_1}{}^{a_2}\wedge \mathfrak{R}_{a_2}{}^{a_1} = -\frac{\mathfrak{R}_2}{(2\pi)^2}\]
where we use the notation
\[\mathfrak{R}_k \equiv \frac{1}{2} \mathfrak{R}_{a_1}{}^{a_2}\wedge \mathfrak{R}_{a_2}{}^{a_3} \ldots  \mathfrak{R}_{a_k}{}^{a_1}\]

Using this, we can write a simple rule to get from the anomaly polynomial to the anomaly-induced
Gibbs free current 
\begin{equation}
\begin{split}
\bar{\mathcal{G}}_{anom} &= u\ \mathcal{P}_{anom}\prn{F\mapsto \mu\ ,\ p_{_1}(\mathfrak{R})\mapsto -T^2} \\
\end{split}
\end{equation}
This elegant result tells us that one can just read off the anomaly-induced Gibbs-free current of 
an ideal Weyl gas in $2d$ directly from its anomaly polynomial. Let us now compare this form with
eqn.\eqref{eqn:Fw} which was derived in \cite{Loganayagam:2011mu} by thermodynamic arguments.
By inspection, it is clear that the above equation follows from \eqref{eqn:Fw} if we take
\begin{equation}
\begin{split}
{\mathfrak{F}}^\omega_{anom} &= \mathcal{P}_{anom}\prn{F\mapsto \mu\ ,\ p_{_1}(\mathfrak{R})\mapsto -T^2} \\
\end{split}
\end{equation}
Hence, there is a straightforward algorithm in $2d$ free fermion theories which takes us 
from the anomaly polynomial of the theory to the polynomial ${\mathfrak{F}}^\omega_{anom}$ 
in $T$ and $\mu$ which determines the anomaly-induced transport.

One of the main aims of the rest of this article is to generalize this result
to arbitrary even dimensions. However, in higher dimensions one does not have
a powerful principle like holomorphy to help us and the anomaly induced 
transport is hence a more subtle effect to derive. So we will spend the next
section to formulate a principle which will help us find the higher 
dimensional analogues of the above result.


\section{Adiabaticity of Spectral flow}\label{sec:adiab}

One of the crucial lessons one draws from the previous section is that there is 
a single Gibbs free energy current from which energy/charge/entropy currents could 
be derived. We will make the reasonable assumption that this continues to 
hold true in higher dimensions. In $2d$ we derived the expression for this
current in terms of a thermal integral
\begin{equation}
\begin{split}
\bar{\mathcal{G}}_{anom} &= \sum_F  \int_0^\infty \frac{dE_p}{2\pi} g_q\ \chi_{_{d=2}}u  \\
\end{split}
\end{equation}
and a natural generalization of the above expression is 
\begin{equation}
\begin{split}
\bar{\mathcal{G}}_{anom}
&= \sum_F \int_0^\infty dE_p\bar{\mathcal{J}}_{q}\ g_q \\
\end{split}
\end{equation}
where $\bar{\mathcal{J}}_{q}$ should be a $2n-1$ form in $d=(2n-1)+1$ dimensions since it
is proportional to the Hodge dual of the Gibbs current 1-form ${\mathcal{G}}^\mu_{anom}$.
Taking a Hodge dual on both sides, we can remove the overbars and write 
\begin{equation}
\begin{split}
{\mathcal{G}}^\mu_{anom}
&= \sum_F \int_0^\infty dE_p{\mathcal{J}}^\mu_{q}\ g_q \\
\end{split}
\end{equation}
Our task is to understand better the physical meaning of $\mathcal{J}^\mu_{q}$ and hence
formulate some sort of an equation for it which then can be solved.

So what is ${\mathcal{J}}^\mu_{q}$ ? It is clear from the way it occurs in the expression for 
Gibbs-current that $\mathcal{J}^\mu_{q}$ is a current whose time-component is just the
1-particle chiral density of states participating in anomalous transport. Hence, we will call it the
\emph{chiral spectral current} from now on. It tells us how a subset of states related to anomaly
flow as the whole fluid(the ideal Weyl gas in this case) flows. In $2d$, we have 
\[ \mathcal{J}^\mu_{q}|_{2d}=\chi_{_{d=2}}\frac{1}{2\pi}\epsilon^{\mu\nu}u_\nu \] 
Hence, two  identical states which differ only in their chirality have opposite  chiral spectral currents.
We will assume that this is true in higher dimensions too $\mathcal{J}^\mu_{q}|_{d=2n}\sim\chi_{_{d=2n}}$
hence justifying the adjective chiral.

In general, we expect the chiral spectral current to be a function of both spacetime co-ordinates $x$ and
the local rest frame energy $E_p$ i.e., $\mathcal{J}^\mu_{q}=\mathcal{J}^\mu_{q}(x,E_p)$
which emphasizes the fact that the chiral spectral current at different energies could be different.
Further, the subscript $q$ denotes the fact that 1-particle states with different charges
can have different chiral spectral currents.

With this understanding, we now turn to the question - how do we determine the chiral spectral
current $\mathcal{J}^\mu_{q}$ for an ideal Weyl gas ? Our crucial tool would be adiabaticity -
as was argued by authors of \cite{Kharzeev:2011ds,Loganayagam:2011mu} (which is a generalization of a
 $d=4$ argument in \cite{Son:2009tf}) the anomaly induced transport in arbitrary dimensions
is heavily constrained by adiabaticity. At the level of states , adiabaticity is basically  
a statement that the states responsible for anomaly induced transport 
do not get created or annihilated\footnote{Except at zero energy due to the anomaly
as we will see below} as they move up or down in the energy or in case of localized
states as they move around in spacetime. This assumption as stated is easily formulated
in terms of a chiral spectral current - it is just the continuity equation for 
the chiral spectral current in the $(x,E_p)$ space,i.e.,
\begin{equation} \nabla_\mu \mathcal{J}^\mu_{q} + \frac{\partial}{\partial E_p} \mathcal{J}^E_{q} =0 \end{equation}  
where $\mathcal{J}^E_{q}(x,E_p)$ is the current of chiral states in the energy direction -
which is the density of states times rate at which their energies are increasing.

For single particle states with the charge $q$ - there are two forces which lead to
a change in energy. First is the rest-frame electric force - the work done by the 
electric force is just the electric field times the charge current 
$E_\mu q\mathcal{J}^\mu_{q}$. Second is the pseudo force (since the rest
frame of the fluid is in general accelerating ) given by $-E_p a_\mu$ where
$a_\mu$ is the acceleration of the fluid. The work done by it
is $-a_\mu$ times the energy current i.e., $-a_\mu E_p\mathcal{J}^\mu_{q}$ .
Combining these together, we get 
\begin{equation}\label{eq:JE} \mathcal{J}^E_{q} =  E_\mu q\mathcal{J}^\mu_{q} - a_\mu E_p\mathcal{J}^\mu_{q} \end{equation}
Taking Hodge duals back again we get the equation that we were after
\begin{equation} 
\begin{split}
D\bar{\mathcal{J}}_{q} + \frac{\partial}{\partial E_p} \bar{\mathcal{J}}^E_{q} =0 \quad\text{with}\quad \bar{\mathcal{J}}^E_{q}  = (qE-E_p a)\wedge \bar{\mathcal{J}}_{q}\\
\end{split}
\end{equation}  
It is easily checked that the $2d$ result $\bar{\mathcal{J}}_{q} = \frac{1}{2\pi}\chi_{_{d=2}} u$ solves the 
above equation. 
 
\section{Chiral spectral current and Anomaly}\label{sec:adiabAnom}

A curious reader might wonder how the equation we just derived relates to the adiabaticity assumption as it 
appears in \cite{Son:2009tf,Kharzeev:2011ds,Loganayagam:2011mu} where the discussion was 
entirely macroscopic with no reference to microscopic states. Further, we have not input
anywhere the information about the anomaly in the above equations. 
In this section we will show that those macroscopic equations 
could be thought of as arising from the above microscopic equation. Further
this would also clarify how anomalies are related to the chiral spectral current.

The flux of density of states is associated with the following energy/charge/entropy currents 
\begin{equation}
\begin{split}
\bar{q}_{anom}&= \sum_F \int_0^\infty dE_p\bar{\mathcal{J}}_{q} E_p f_q \\
\bar{J}_{anom}&= \sum_F \int_0^\infty dE_p\bar{\mathcal{J}}_{q} q f_q \\
\bar{J}_{S,anom}&= -\sum_F \int_0^\infty dE_p\bar{\mathcal{J}}_{q}\mathcal{H}_q\\ 
\end{split}
\end{equation}

For a general system with anomalies, the statement that these transport processes
have to be adiabatic is equivalent to the following equation (as shown in \cite{Loganayagam:2011mu}) 
\begin{equation}\label{eq:hydroAdiab}
\begin{split}
D\bar{q}_{anom}+a\wedge \bar{q}_{anom} - E\wedge \bar{J}_{anom} &= T D \bar{J}_{S,anom} + \mu \brk{ D \bar{J}_{anom} -\bar{\mathfrak{A}}[F]}\\
\end{split}
\end{equation}
where $\bar{\mathfrak{A}}[F]$ is the anomaly $2n$-form in a given theory.   This equation above assumes that one is working in flat spacetime and hence all gravitational anomalies are turned off. We will refer the reader 
to \cite{Loganayagam:2011mu} for a proper derivation of this equation and its consequences for
a general system. Here we are mainly interested in exploring what it means for the chiral
spectral current. Substituting the above expressions for the various currents and using the following relations 
\begin{equation}
\begin{split}
f_q \equiv \frac{1}{e^{\beta(E_p-q\mu)}+1} &= \frac{\partial g_q}{\partial E_p}\\
E_pf_q+T\mathcal{H}_q-q\mu f_q &= g_q\\
E_p Df_q+TD\mathcal{H}_q-q\mu Df_q &= 0\\
\end{split}
\end{equation}
the adiabaticity equation of \cite{Loganayagam:2011mu} assumes an especially simple form
\begin{equation}
\begin{split}
\sum_F \int_0^\infty dE_p\brk{g_q D\bar{\mathcal{J}}_q  -\bar{\mathcal{J}}^E_q \frac{\partial g_q}{\partial E_p} } +\mu \bar{\mathfrak{A}} &=0\\
\end{split}
\end{equation}
where as before $\bar{\mathcal{J}}^E_{q}  = (qE-E_p a)\wedge \bar{\mathcal{J}}_{q}$. In this form, the relation to the 
microscopic equation that we had derived before is evident. To see this we should integrate by parts - this should
be done carefully since the boundary contributions do not vanish. We get
\begin{equation}
\begin{split}
\sum_F \int_0^\infty dE_p\ g_q\brk{ D\bar{\mathcal{J}}_q  + \frac{\partial}{\partial E_p}\bar{\mathcal{J}}^E_q } +\mu \bar{\mathfrak{A}} -\sum_F \brk{g_q\bar{\mathcal{J}}^E_q}_{E_p=0}^{E_p=\infty} &=0\\
\end{split}
\end{equation}
We can now use the continuity equation for the chiral spectral current to set the integrand inside
the integral to zero.

Now we turn to the boundary contributions : first the UV contribution -  
 $g_q$ falls exponentially as the rest-frame energy $E_p\to \infty$ and it is 
reasonable to assume that the growth of $\bar{\mathcal{J}}^E_q $ is slow enough 
that the contribution from  $E_p= \infty$ is zero \footnote{In fact, 
if $\bar{\mathcal{J}}^E_q $ did grow exponentially with energy the various integrals
we have been writing down would all be UV divergent and we would have to worry how
to make sense out of them. We will assume that this is not the case and this assumption can 
be justified in explicit examples.}. This is consistent with the fact that anomalies and their 
related transport processes are not UV-sensitive despite the fact that historically 
they were discovered in calculations where one should renormalize UV-sensitive quantities.

Now we turn to the IR contribution - there is no good reason for the boundary contribution 
from $E_p=0$ to vanish and in fact we need it to balance the contribution from the anomaly. 
Equating the anomaly to this IR contribution, we get 
\begin{equation}\label{eq:AnomJE}
\begin{split}
\bar{\mathfrak{A}} =-\frac{1}{\mu}\sum_F g_q(E_p=0)\ \bar{\mathcal{J}}^E_q(E_p=0) \\
\end{split}
\end{equation}
This is an interesting equation which is a version of the well-known relation between
the spectral flow of chiral zero energy states and the  anomaly in the system.
To see how this might work,  we will now write the anomaly term also as a sum.
For a set of charged Weyl fermions in $d=(2n-1)+1$ dimensional flat spacetime, 
the covariant anomaly is given by
\begin{equation}
\begin{split} 
\bar{\mathfrak{A}}[F] = -\frac{1}{n!}\prn{\frac{F}{2\pi}}^n\sum_{species} \chi_{_{d=2n}} q^{n+1}
\end{split}
\end{equation}
where the sum runs over every species with one term appearing in the sum for every particle-antiparticle
couple \footnote{It does not matter which among that couple is chosen, since $\chi_{_{d=2n}} q^{n+1}$ is the
same for both - this follows from the fact that if the particles charge and chirality is $(q,\chi_{_{d=2n}})$
then the antiparticle's charge/chirality  is $(-q,(-1)^{n-1}\chi_{_{d=2n}})$ .}. 

Let us divide the field-strength $F$ appearing in the above equations into electric and magnetic
fields in the rest frame of the fluid. The rest frame electric field is 
$E_\mu = u^\nu F_{\mu\nu}$  which we can think of as a 1-form $E$. The rest frame magnetic field
is a 2-form obtained by subtracting the electric part from $F$, 
\[ B\equiv F-u\wedge E \]
where $u=u_\mu dx^\mu$ is the velocity 1-form. Substituting $F=B-E\wedge u$ into the anomaly above we get
\begin{equation}
\begin{split}
\mu\bar{\mathfrak{A}}[F]
&=\sum_{species}\chi_{_{d=2n}} qE\wedge \frac{q\mu}{2\pi} \prn{\frac{qB}{2\pi}}^{n-1}\wedge \frac{u}{(n-1)!}
\end{split}
\end{equation}
Now use the fact that $g_q(E_p=0)-g_{-q}(E_p=0)= -q\mu$ to get 
\begin{equation}
\begin{split}
\bar{\mathfrak{A}}[F]&= -\frac{1}{\mu}\sum_F g_q(E_p=0)\chi_{_{d=2n}}  \frac{qE}{2\pi} \wedge\prn{\frac{qB}{2\pi}}^{n-1}\wedge \frac{u}{(n-1)!}
\end{split}
\end{equation}
where the sum now runs over all the fermions with particles and antiparticles counted separately. 
Comparing this with \eqref{eq:AnomJE}, we get for each particle
\[\bar{\mathcal{J}}^E_q(E_p=0) = qE\wedge\bar{\mathcal{J}}_q(E_p=0) = \frac{qE}{2\pi}  \wedge\chi_{_{d=2n}}  \prn{\frac{qB}{2\pi}}^{n-1}\wedge \frac{u}{(n-1)!} \] 
or 
\begin{equation}
\bar{\mathcal{J}}_q(E_p=0) = \frac{\chi_{_{d=2n}}}{2\pi}    \prn{\frac{qB}{2\pi}}^{n-1}\wedge \frac{u}{(n-1)!}
\end{equation}
This flow of density of states at zero energy has a direct explanation in terms of Landau level physics
in the local rest frame of the fluid. We review this connection in more detail in 
the appendix\S\ref{app:magnetic} and relate it to the chiral magnetic effect in ideal Weyl gas.
Hence, to conclude the following equations
\begin{equation}\label{eq:adiabMicro}
\begin{split}
D\bar{\mathcal{J}}_{q} &+ \frac{\partial}{\partial E_p} \bar{\mathcal{J}}^E_{q} =0 \quad\text{with}\\\
\bar{\mathcal{J}}^E_{q}  &= (qE-E_p a)\wedge \bar{\mathcal{J}}_{q}\\
\bar{\mathcal{J}}_q(E_p=0) &= \frac{\chi_{_{d=2n}}}{2\pi}    \prn{\frac{qB}{2\pi}}^{n-1}\wedge \frac{u}{(n-1)!}\\
\end{split}
\end{equation}
implies for an ideal Weyl gas the macroscopic adiabaticity condition in eqn.\eqref{eq:hydroAdiab}.
Our next task is to solve these equations for an arbitrary fluid 
flow which we will do in the next section.

\section{Solving the adiabaticity equation}\label{sec:solvAdiab}
We now seek a solution of eqn.\eqref{eq:adiabMicro} for an arbitrary fluid flow of an ideal
Weyl gas. We begin by  decomposing the exterior derivative of the velocity
into its vorticity part and the acceleration part
\begin{equation} Du = 2\omega -u\wedge a \end{equation}
where $\omega$ is the vorticity 2-form with $\omega_{\mu\nu}u^\nu=0$. We
will now make an ansatz for the solution of the form
\begin{equation}\label{eq:ansatz}
\begin{split}
\bar{\mathcal{J}}_q = \frac{\chi_{_{d=2n}}}{(2\pi)^n} \sum_{k=0}^{n-1} \alpha_k \frac{(2\omega E_p)^k}{k!}\wedge \frac{(qB)^{n-1-k}}{(n-1-k)!} \wedge u
\end{split}
\end{equation}
This ansatz is motivated by the solution to the macroscopic adiabaticity equations presented in 
\cite{Loganayagam:2011mu} which took the form of a polynomial in 2-forms $B$ and $\omega$. The
powers of $E_p$ are then fixed by dimensional analysis. The numerical coefficients are
arranged such that the $E_p=0$ boundary condition fixes $\alpha_0=1$. We now 
want to substitute this ansatz into the eqn.\eqref{eq:hydroAdiab} to fix other 
$\alpha_k\ $s. Using the following identities\cite{Loganayagam:2011mu}
\begin{equation}
\begin{split}
Du &= 2\omega +a\wedge u \\
D(qB)\wedge u &= -qE\wedge 2\omega \wedge u \\
D(2\omega)\wedge u&= a\wedge 2\omega\wedge u\\
\end{split}
\end{equation}
we get
\begin{equation}
\begin{split}
D\bar{\mathcal{J}}_q&= -\frac{\chi_{_{d=2n}}}{(2\pi)^n} \sum_{k=0}^{n-1}\brk{ k\alpha_{k-1} \frac{qE}{E_p}-(k+1)\alpha_k a}\wedge \frac{(2\omega E_p)^k}{k!}\wedge \frac{(qB)^{n-1-k}}{(n-1-k)!} \wedge u\\
\end{split}
\end{equation}
where we have in addition used the fact that any $2n$ form made of purely spatial forms $B$ and $\omega$ is zero.
On the other hand 
\begin{equation}
\begin{split}
\frac{\partial}{\partial E_p}\bar{\mathcal{J}}^E_q&= \frac{\chi_{_{d=2n}}}{(2\pi)^n} \sum_{k=0}^{n-1}\brk{ k\alpha_k \frac{qE}{E_p}-(k+1)\alpha_k a}\wedge \frac{(2\omega E_p)^k}{k!}\wedge \frac{(qB)^{n-1-k}}{(n-1-k)!} \wedge u\\
\end{split}
\end{equation}
and demanding that the sum of the last two equations should vanish sets $\alpha_k=\alpha_{k-1}$ for all $k\geq 1$.
Along with the boundary condition at $E_p=0$ which sets $\alpha_0=1$ this determines $\alpha_k=1$ for all k.
Substituting this into our ansatz, we finally get
\begin{equation}\label{eq:soln}
\begin{split}
\bar{\mathcal{J}}_q &= \frac{\chi_{_{d=2n}}}{2\pi}    \prn{\frac{qB+2\omega E_p}{2\pi}}^{n-1}\wedge \frac{u}{(n-1)!}\\
\end{split}
\end{equation}
This expression is the central result of the article - it is a formula for how the chiral states of 
a given energy flow when the fluid flows. Since we have not invoked any equations of motion for 
the fluid in our derivation, this is an `off-shell' solution valid for arbitrary fluid flows
- a microscopic analogue  of the off-shell solution derived in \cite{Loganayagam:2011mu}. In the
rest of the article we will explore various  consequences of the above formula.

\section{Anomaly/transport in Ideal Weyl gas of arbitrary dimensions}\label{sec:FwSoln}
We begin by substituting for $\bar{\mathcal{J}}_{q}$ in the expression for $\bar{\mathcal{G}}_{anom}$
\begin{equation}
\begin{split}
\bar{\mathcal{G}}_{anom} &= \sum_F \int_0^\infty dE_p\bar{\mathcal{J}}_{q}\ g_q \\
&= \sum_F \chi_{_{d=2n}}\int_0^\infty \frac{dE_p }{2\pi}    \prn{\frac{qB+2\omega E_p}{2\pi}}^{n-1}\wedge \frac{u}{(n-1)!}\ g_q \\
\end{split}
\end{equation}
To evaluate this integral, we will again pair the particles and anti-particles together and use the 
fact that if the charge/chirality  of a particle is $(q,\chi_{_{d=2n}})$ then the charge/chirality
of the anti-particle is $(-q,(-1)^{n-1}\chi_{_{d=2n}})$.
\begin{equation}
\begin{split}
&\bar{\mathcal{G}}_{anom}\\
&= \sum_{species} \int_0^\infty \frac{dE_p }{2\pi}  \brk{ g_q \prn{\frac{qB+2\omega E_p}{2\pi}}^{n-1}+(-1)^{n-1}g_{-q} \prn{\frac{-qB+2\omega E_p}{2\pi}}^{n-1} } \wedge \frac{\chi_{_{d=2n}} u}{(n-1)!}\  \\
&= \sum_{species} \int_0^\infty \frac{dE_p }{2\pi}  \brk{ g_q \prn{\frac{qB+2\omega E_p}{2\pi}}^{n-1}+g_{-q} \prn{\frac{qB-2\omega E_p}{2\pi}}^{n-1} } \wedge \frac{\chi_{_{d=2n}} u}{(n-1)!}\  \\
\end{split}
\end{equation}
To proceed further, it is convenient to employ a formal trick - the trick is to construct a 
generating function which will in one sweep contain in its Taylor expansion
$\bar{\mathcal{G}}_{anom}$ of arbitrary even dimensions. To this end let us multiply
the above expression by $\tau^{n-1}$ where $\tau$ is a formal parameter and perform
a sum over all integers $n\geq 1$. We get
\begin{equation}
\begin{split}
&\sum_{n=1}^{\infty}\tau^{n-1}\prn{\bar{\mathcal{G}}_{anom}}_{d=2n}\\
&= \sum_{n=1}^{\infty}\sum_{species} \int_0^\infty \frac{dE_p }{2\pi}  \brk{ g_q \prn{\frac{qB+2\omega E_p}{2\pi}}^{n-1}+g_{-q} \prn{\frac{qB-2\omega E_p}{2\pi}}^{n-1} } \wedge \frac{\tau^{n-1}\chi_{_{d=2n}} u}{(n-1)!}\  \\
&=\sum_{species} e^{\frac{\tau}{2\pi}qB} \int_0^\infty \frac{dE_p }{2\pi}  \brk{ g_q e^{\frac{\tau}{2\pi}2\omega E_p} +g_{-q}  e^{-\frac{\tau}{2\pi}2\omega E_p} } \wedge \chi_{_{d}} u\  \\
\end{split}
\end{equation}
where we have used $\chi_{_{d}}$ to represent the chirality in the appropriate dimension.

We evaluate this integral in the Appendix\S\ref{app:fermifn}. The final result is 
\begin{equation}\label{eq:GEval}
\begin{split}
&\sum_{n=1}^{\infty}\tau^{n-1}\prn{\bar{\mathcal{G}}_{anom}}_{d=2n}\\
&=-\sum_{species} e^{\frac{\tau}{2\pi}qB} \frac{2\pi}{(2\omega\tau)^2}\brk{\frac{\frac{\tau}{2}2\omega\beta^{-1}}{\sin \frac{\tau}{2}2\omega\beta^{-1}}e^{\frac{\tau}{2\pi}2\omega q\mu}-\prn{1+\frac{\tau}{2\pi}2\omega q\mu}} \wedge \chi_{_{d}} u\ \\
\end{split}
\end{equation}
or 
\begin{equation}
\begin{split}
&\sum_{n=1}^{\infty}\tau^{n+1}\prn{\bar{\mathcal{G}}_{anom}}_{d=2n}\\
&=- \frac{1}{(2\omega)^2}\sum_{species}\brk{2\pi\frac{\frac{\tau}{2}2\omega\beta^{-1}}{\sin \frac{\tau}{2}2\omega\beta^{-1}}e^{\frac{q\tau}{2\pi}(B+2\omega\mu)}-2\pi e^{\frac{\tau}{2\pi}qB} \prn{1+\frac{\tau}{2\pi}2\omega q\mu}} \wedge \chi_{_{d}} u\ \\
\end{split}
\end{equation}

We want to now compare this with the form eqn\eqref{eqn:Fw} derived in \cite{Loganayagam:2011mu} which 
we reproduce for the convenience of the reader
\begin{equation}
\begin{split}
\bar{\mathcal{G}}_{anom}
&=\frac{1}{(2\omega)^2}\Bigl\{\mathfrak{F}^\omega_{anom}[T(2\omega),B+\mu(2\omega)]-\Bigl[\mathfrak{F}^\omega_{anom}[T(2\omega),B+\mu(2\omega)]\Bigr]_{\omega=0}
 \Bigr.\\
&\qquad\qquad\qquad\Bigl.-\quad\omega\Bigl[\frac{\delta}{\delta\omega}\mathfrak{F}^\omega_{anom}[T(2\omega),B+\mu(2\omega)]\Bigr]_{\omega=0} \Bigr\}\wedge u\\ 
\end{split}
\end{equation}
By comparing term by term, we get a simple expression
\begin{equation}
\begin{split}
\prn{{\mathfrak{F}}^\omega_{anom}}_{d=2n}&=- 2\pi\sum_{species}  \chi_{_{d=2n}} \brk{\frac{\frac{\tau}{2}T}{\sin \frac{\tau}{2}T}e^{\frac{\tau}{2\pi}q\mu}}_{\tau^{n+1}} \\
\end{split}
\end{equation}
where the subscript $\tau^{n+1}$ denotes that one needs to Taylor-expand in $\tau$ and
retain the coefficient of $\tau^{n+1}$.

Our aim was to give a prescription to get to ${\mathfrak{F}}^\omega_{anom}$ from the corresponding 
anomaly polynomial.  The anomaly polynomial of an Ideal Weyl gas is  given by (see appendix \ref{app:anom} 
for explicit expressions in various dimensions)
\[ \prn{\mathcal{P}_{anom}}_{d=2n}=- 2\pi\sum_{species}  \chi_{_{d=2n}}\ \brk{\hat{A}\prn{\tau \mathfrak{R}}\ 
e^{\frac{\tau}{2\pi}qF}}_{\tau^{n+1}}\]
Now using (see appendix \ref{app:anom} for a derivation)
\begin{equation}
\brk{\hat{A}\prn{\tau \mathfrak{R}}}_{ p_{_{k>1}}(\mathfrak{R}) =  0} = \frac{\frac{\tau}{2}\sqrt{-p_{_1}(\mathfrak{R})}}{\sin\prn{\frac{\tau}{2}\sqrt{-p_{_1}(\mathfrak{R})}}}
\end{equation}
we can write 
\begin{equation}\label{eq:subs}
\begin{split}
{\mathfrak{F}}^\omega_{anom}&=\mathcal{P}_{anom}\prn{F\mapsto \mu ,\ p_{_1}(\mathfrak{R}) \mapsto -T^2,\ p_{_{k>1}}(\mathfrak{R}) \mapsto  0 }
\end{split}
\end{equation}
in any dimension for arbitrary collection of free fermions. This is a remarkably elegant result which
takes the anomaly polynomial and via simple substitutions converts it into the polynomial which 
governs the anomaly-induced transport.

\section{Discussion}\label{sec:disc}

One of our main motivations in undertaking this study was to understand at a more microscopic
level how the anomaly-induced transport comes  about. While the transport proportional 
to the magnetic field can be understood relatively easily\footnote{By a Landau level
argument which we review in appendix \S\ref{app:magnetic}}, the microscopic origins of 
the vortical effect were relatively obscure. Having understood now the relevant 
microscopic dynamics as that of spectral flow, one might ask whether one can proceed
away from equilibrium. In other words, how does this picture of chiral spectral
current help us understand the non-equilibrium phenomena driven by anomaly.

While we do not have a complete answer to this question, let us give an idea of what the
form the answer might take. Ideally to answer non-equilibrium questions, one would 
like to have a Boltzmann-type equation describing an ideal Weyl gas to which interactions
can be added via collision terms. The Boltzmann equation is intimately tied to the flow
of states in the semi-classical phase-space : in fact, it is just a conservation-type equation on the
phase space. Let $\zeta^A$ be the co-ordinates on the phase-space and $\mathcal{J}^A_{tot}$ be the
net flow of states (the total spectral current) in the phase-space. Then the 
Boltzmann equation takes the schematic form 
\[ \frac{\partial}{\partial \zeta^A}\brk{f \mathcal{J}^A_{tot}} = C[f] \]
where the RHS is the collision term which takes into account the scattering from one point
in the phase-space to the other point. Assuming the total number of states is
conserved, we can assume $\partial_A \mathcal{J}^A_{tot}  = 0$
so that the Boltzmann equation reduces to 
\[ \mathcal{J}^A_{tot}\partial_A f = C[f] \]
Hence, we see that a spectral current in the phase space leads us immediately to a Boltzmann
type equation. To proceed further we need to figure out what the expression for $\mathcal{J}^A_{tot}$
is ? Conventional kinetic theory would suggest that 
\begin{equation}
\begin{split}
\dot{x}^\mu &\sim p^\mu \quad\text{and}\quad  \dot{p}_\mu \sim q p^\nu F_{\mu\nu}\\
\implies \mathcal{J}^A_{naive}\partial_A  &\sim p^\mu\frac{\partial}{\partial x^\mu} + q p^\nu F_{\mu\nu} \frac{\partial}{\partial p_\mu} 
\end{split}
\end{equation}
which would be identical to the Boltzmann operator one would write down for a Dirac fermion. 
But, as we had argued in this article, in an ideal Weyl gas there is an \emph{additional}
contribution to the spectral current over and above the contribution in a Dirac gas. This 
should lead to the modification of the naive Boltzmann operator that we had written above.
Unfortunately, the spectral current that we have written down in this article has no 
momentum information - it is only a function of $\{x^\mu,E_p\}$ and hence we cannot 
yet write down the exact modification of the Boltzmann operator. It would be nice to 
work out the momentum-resolved chiral spectral current in order to write down this modification .

We will now give an alternate argument on why we expect such a modification to the Boltzmann
equation - as is known in the phenomenology of Weyl semi-metals\cite{2011PhRvB..83t5101W,
2011PhRvB..84g5129Y} (these are $3+1$d phases with Weyl fermionic excitations ), 
the Weyl fermion in $3+1$d should be thought of as a source of Berry flux 
in the momentum space. The presence of such a Berry-flux is known to lead to
exactly the kind of modifications of the Boltzmann equations \cite{2004PhRvL..93t6602H,
2010RvMP...82.1959X} that we propose above. It would be nice to clarify the relation
between such Berry phase related ideas and the ideas presented in this article.

Let us recap : what we have done is to use adiabaticity as the basic idea which drives the 
anomaly-induced transport in Ideal Weyl gases. This has resulted in a simple rule given by
eqn.\eqref{eq:subs} which determines completely the polynomial introduced in \cite{Loganayagam:2011mu}.
The novel result is the way gravitational anomalies seem to enter into the temperature 
dependence as was noted by \cite{Landsteiner:2011iq}. While we have reproduced their $4d$ results,
we have done it by an entirely different method - they had employed the $4d$ Kubo formulae derived
in \cite{Amado:2011zx} whereas our derivation relies on adiabaticity.

Kubo formulae for anomaly-induced transport in higher dimensions involve higher point functions
and since the Kubo formalism is not well-developed beyond two-point functions, it is \emph{a priori}
more difficult to take that route for calculations in higher dimensions. In this article, rather
than generalize the Kubo formulae in \cite{Amado:2011zx} we have taken an alternate route. But 
having got the answer, let us provide a guess as to how the results of \cite{Amado:2011zx} would
generalize. Consider the small frequency/small momentum limit of the following $n$-point
function in $d=2n$ dimensions at finite $T,\mu$ :
\begin{equation}
\begin{split}
&\epsilon^{0i_1 i_2\ldots i_n j_1 j_2\ldots j_{n-1}} 
\langle\ T_{0i_1}(k_1^\alpha) T_{0i_2}(k_2^\alpha)\ldots T_{0i_l}(k_l^\alpha)
J_{i_{l+1}}(k_{l+1}^\alpha) J_{i_{l+2}}(k_{l+2}^\alpha) \ldots J_{i_n}(k_n^\alpha)\ \rangle \\
&\qquad \sim \xi_{l}(T,\mu)\  \delta^{2n}\prn{k_1+k_2+\ldots+k_n} k_1^{[j_1}k_2^{j_2}\ldots k_{n-1}^{j_{n-1}]} 
\end{split}
\end{equation}
then generalizing the results of \cite{Amado:2011zx},  it is tempting to conjecture 
that $\xi_{l}(T,\mu)$ is the transport coefficient associated with $(2\omega)^{l-1}\wedge B^{n-l}\wedge u$
term in $\bar{q}_{anom}$ or equivalently  $(2\omega)^{l}\wedge B^{n-l-1}\wedge u$
term in $\bar{J}_{anom}$ since these two are related by the generalised Onsager type relation \cite{Loganayagam:2011mu}
\begin{equation}
\begin{split}
\frac{\delta \bar{q}_{anom}}{\delta B} = \frac{\delta \bar{J}_{anom}}{\delta (2\omega)}
\end{split}
\end{equation}
This in particular would mean that for a free theory we expect the transport coefficient $\xi_{l}(T,\mu)$
to be related to a Fermi-Dirac integral of the type \footnote{We would like to thank Carlos Hoyos for suggesting
such a relation.} 
\[ \xi_{l}(T,\mu) \sim \int_0^\infty \frac{dE_p}{2\pi} E_p^l\brk{f_q-(-1)^l f_{-q}}\]  
It would be nice to show that such a Kubo-type formula as conjectured above does hold\footnote{This would presumably involve
generalising the Kubo formalism to higher point functions\cite{Chou:1984es,
Moore:2010bu,Barnes:2010jp,Arnold:2011ja,Saremi:2011nh,Arnold:2011hp}. }, since it then opens up the possibility 
that we can calculate such transport coefficients in any field theory. In particular, such a
Kubo formula would be of enormous use in checking whether the substitution rule in
eqn.\eqref{eq:subs} holds beyond free theory. It would be interesting to see what happens
to this rule as the effective anomaly polynomial is changed via Green-Schwarz mechanism.

There is already some evidence that this rule holds in strongly coupled holographic phases
\cite{Landsteiner:2011cp}. It would be interesting to generalize this by 
constructing  and studying rotating black hole solutions in AdS$_{2n+1}$ with pure/mixed 
gravitational Chern-Simons terms. This would presumably link the  proposed modification of Wald entropy
\cite{Tachikawa:2006sz,Bonora:2011gz} in the presence of such Chern-Simons terms
to the anomaly-induced entropy transport in the dual CFT.  

We do not yet have an intuitive field theory understanding why the higher
Pontryagin classes do not contribute to the anomaly induced transport that we discuss in this article.
One way to undertsand this statement better would be to see whether this is true in 
fluids with gravity duals via the calculation we just outlined. One way this might work out 
is if higher Pontryagin classes fall off too fast in the bulk to contribute to 
boundary transport while the first Pontryagin class falls off slowly and asymptotes to  
$- T^2 (2\omega)^2$ near the AdS boundary \footnote{We would like to thank 
Karl Landsteiner and Mukund Rangamani for discussions regarding this issue.}.

Another way in which the $T^2$ coefficients in $4d$ were calculated was by Vilenkin using
rotating ensembles \cite{Vilenkin:1978hb,Vilenkin:1979ui,Vilenkin:1980fu,Vilenkin:1980zv,Vilenkin:1980ft,Vilenkin:1995um}.
This method as Vilenkin used is plagued with technical subtleties which have to do with
trying to define rigid rotation in a relativistic field theory in the Minkowski
spacetime. But these subtleties can be avoided by studying rotating ensembles on 
a \emph{sphere} which serves as an infrared regulator. We have done some preliminary
calculations in Weyl fermions using this method and they seem to agree with the results
presented here. The details of this calculation will be presented elsewhere. 

Further it would be interesting to generalise all these arguments to spacetimes with
non-zero curvature. A first step towards such a generalisation would be to first classify
the various curvature dependent transport processes which are sensitive to anomalies.
Since a free theory of Weyl fermions is conformal, we can restrict our attention to 
just the Weyl-covariatised curvature tensors\cite{Loganayagam:2008is} and the holographic
computations that we had outlined above could be easily used to identify the transport processes
which are sensitive to anomalies. It would be satisfying to understand 
the chiral spectral current in the strongly interacting systems via holography \footnote{We would
like to thank Jyotirmoy Bhattacharya for discussions regarding this issue}.  

\subsection*{Acknowledgements}
It is a pleasure to thank  Tatsuo Azeyanagi, Koushik Balasubramanian,
Jyotirmoy Bhattacharya, Sayantani Bhattacharyya, Alejandra Castro,
Amar V. Chandra, Clay Cordova, Thomas Faulkner, Liang Fu, Alessandra Gnecchi, Andrzej G\"{o}rlich,
Tarun Grover, Sean Hartnoll, Carlos Hoyos, Nabil Iqbal, Romuald Janik, Matthias Kaminski, Vijay Kumar, 
Karl Landsteiner, Tongyan Lin,
Hong Liu, Andreas W. W. Ludwig, Juan Maldacena, John McGreevy, Max Metlitski,
Shiraz Minwalla, Gim-Seng Ng, Mukund Rangamani, Arnab Sen, David Simmons-Duffin,
Tarun Sharma, Dam Thanh Son, Wei Song, Andrew Strominger, Washington Taylor, Dan Thompson,  
Senthil Todadri and Naoki Yamamoto for various useful discussions on ideas presented in this 
paper. RL would like to thank  \textbf{
Conference on Cold Materials, Hot Nuclei and Black Holes:  Applied Gauge/Gravity Duality}
at the Abdus Salam International Centre for Theoretical Physics , Trieste and
\textbf{Holographic Duality and Condensed Matter Physics workshop} at the Kavli Institute
for Theoretical Physics, UCSB, Santa Barbara for their hospitality while this
work was being completed. PS would like to thank Harvard University, Institute for Nuclear Theory in Seattle and Niels Bohr Institute in Copenhagen for hospitality while this work was completed. RL is supported by the Harvard Society of Fellows 
through a junior fellowship. PS is supported in part by the Belgian Federal Science Policy Office through
the Interuniversity Attraction Pole IAP VI/11 and by FWO-Vlaanderen through project
G011410N. Finally, RL would  like to thank various colleagues at the 
Harvard society for interesting discussions. 

\section*{Appendices}
\appendix

\section{Notes on Fermi-Dirac functions}\label{app:fermifn}
Our objective in this section is to evaluate various moments 
associated with Fermi-Dirac distributions. We will start by defining 
some basic functions which would be useful later on.
First  we want to define the $n^{th}$ poly-logarithm $Li_n(x)$ defined via 
\begin{equation}
\begin{split}
Li_n(x) &\equiv \sum_{k=1}^\infty \frac{x^k}{k^n} =\sum_{k=1}^\infty \int_0^\infty dy (xe^{-y})^k  \frac{y^{n-1}}{(n-1)!}\\
&=\int_0^\infty dy\ \frac{1}{x^{-1}e^y-1}\ \frac{y^{n-1}}{(n-1)!}=\int_0^\infty dy\ \frac{d}{dy}\brk{\ln\prn{1-xe^{-y}}}\ \frac{y^{n-1}}{(n-1)!}\\
&=\int_0^\infty dy\ \brk{-\ln\prn{1-xe^{-y}}}\ \frac{y^{n-2}}{(n-2)!}\\
\end{split}
\end{equation}
the last line being valid for $n\geq 2$. The importance of polylogarithm
lies in its relation to various moments of Fermi-Dirac integrals - it follows from above that 
\begin{equation}
\begin{split}
\int_0^\infty dE_p \frac{E_p^k}{k!} g_q &= \frac{1}{\beta^{k+2}}Li_{k+2}\prn{-e^{\beta q\mu}}\\
\int_0^\infty dE_p \frac{E_p^k}{k!} f_q &= -\frac{1}{\beta^{k+1}}Li_{k+1}\prn{-e^{\beta q\mu}}\\
\end{split}
\end{equation}

We introduce next  the $n^{th}$ Bernoulli Polynomials  $B_n(x)$ - these are polynomials 
defined via the following generating function
\begin{equation}\label{eq:BernoulliGen}
\begin{split}
\frac{t\ e^{xt}}{e^t-1} \equiv \sum_{n=0}^\infty  \frac{t^n}{n!}\ B_n(x)
\end{split}
\end{equation}
We now write the identity relating poly-logarithms with Bernoulli polynomials 
(which itself is a special case of the relation between poly-logarithms and 
Hurwitz zeta function)
\begin{equation}
\begin{split}
Li_{n}\prn{-e^{\beta q\mu}} +(-1)^{n}Li_{n}\prn{-e^{-\beta q\mu}} &=- \frac{(2\pi i)^{n}}{n!}B_{n}\prn{\frac{1}{2}+\frac{\beta q\mu}{2\pi i}}
\end{split}
\end{equation}
which implies the following identity for the Fermi-Dirac moments
\begin{equation}\label{eq:gPoly}
\begin{split}
\int_0^\infty \frac{dE_p}{2\pi} \frac{1}{k!}\prn{\frac{E_p}{2\pi}}^k \brk{g_q+(-1)^k g_{-q}} &= -\frac{2\pi}{(k+2)!}\prn{\frac{ i}{\beta}}^{k+2} B_{k+2}\prn{\frac{1}{2}+\frac{\beta q\mu}{2\pi i}}\\
\end{split}
\end{equation}
or more explicitly
\begin{equation}\label{eq:gPolyExp}
\begin{split}
\int_0^\infty \frac{dE_p}{2\pi}  \brk{g_q+ g_{-q}} &= -2\pi \brk{\frac{1}{2!}\prn{\frac{q\mu}{2\pi}}^2+\frac{T^2}{4!} }\\
\int_0^\infty \frac{dE_p}{2\pi}  \prn{\frac{E_p}{2\pi}} \brk{g_q- g_{-q}} &= -2\pi \brk{\frac{1}{3!}\prn{\frac{q\mu}{2\pi}}^3+ 
 \prn{\frac{q\mu}{2\pi}}\frac{T^2}{4!}  } \\
\int_0^\infty \frac{dE_p}{2\pi}  \prn{\frac{E_p}{2\pi}}^2 \brk{g_q + g_{-q}} &= -2\pi \brk{\frac{1}{4!}\prn{\frac{q\mu}{2\pi}}^4+\frac{1}{2!}\prn{\frac{q\mu}{2\pi}}^2 \frac{ T^2}{4!}+\frac{7}{8}\frac{T^4}{6!}}\\
\int_0^\infty \frac{dE_p}{2\pi}  \prn{\frac{E_p}{2\pi}}^3 \brk{g_q - g_{-q}} &= -2\pi \brk{\frac{1}{5!}\prn{\frac{q\mu}{2\pi}}^5+\frac{1}{3!}\prn{\frac{q\mu}{2\pi}}^3 \frac{ T^2}{4!}+\prn{\frac{q\mu}{2\pi}} \frac{7}{8}\frac{T^4}{6!}}\\
\int_0^\infty \frac{dE_p}{2\pi}  \prn{\frac{E_p}{2\pi}}^4 \brk{g_q + g_{-q}} &= -2\pi \brk{\frac{1}{6!}\prn{\frac{q\mu}{2\pi}}^6+\frac{1}{4!}\prn{\frac{q\mu}{2\pi}}^4 \frac{ T^2}{4!}+\frac{1}{2!}\prn{\frac{q\mu}{2\pi}}^2 \frac{7}{8}\frac{T^4}{6!}+\frac{31}{24}\frac{T^6}{8!}}\\
\end{split}
\end{equation}

Similarly, we can write 
\begin{equation}\label{eq:fPoly}
\begin{split}
\int_0^\infty \frac{dE_p}{2\pi} \frac{1}{k!}\prn{\frac{E_p}{2\pi}}^k \brk{f_q-(-1)^k f_{-q}} &= \frac{1}{(k+1)!}\prn{\frac{ i}{\beta}}^{k+1} B_{k+1}\prn{\frac{1}{2}+\frac{\beta q\mu}{2\pi i}}\\
\end{split}
\end{equation}
or more explicitly
\begin{equation}\label{eq:fPolyExp}
\begin{split}
\int_0^\infty \frac{dE_p}{2\pi}  \brk{f_q- f_{-q}} &=\prn{\frac{q\mu}{2\pi}}\\
\int_0^\infty \frac{dE_p}{2\pi}  \prn{\frac{E_p}{2\pi}} \brk{f_q+ f_{-q}} &=\frac{1}{2!}\prn{\frac{q\mu}{2\pi}}^2+\frac{T^2}{4!} \\
\int_0^\infty \frac{dE_p}{2\pi}  \prn{\frac{E_p}{2\pi}}^2 \brk{f_q- f_{-q}} &= {\frac{1}{3!}\prn{\frac{q\mu}{2\pi}}^3+ 
 \prn{\frac{q\mu}{2\pi}}\frac{T^2}{4!}  } \\
\int_0^\infty \frac{dE_p}{2\pi}  \prn{\frac{E_p}{2\pi}}^3 \brk{f_q + f_{-q}} &= {\frac{1}{4!}\prn{\frac{q\mu}{2\pi}}^4+\frac{1}{2!}\prn{\frac{q\mu}{2\pi}}^2 \frac{ T^2}{4!}+\frac{7}{8}\frac{T^4}{6!}}\\
\int_0^\infty \frac{dE_p}{2\pi}  \prn{\frac{E_p}{2\pi}}^4 \brk{f_q - f_{-q}} &= {\frac{1}{5!}\prn{\frac{q\mu}{2\pi}}^5+\frac{1}{3!}\prn{\frac{q\mu}{2\pi}}^3 \frac{ T^2}{4!}+\prn{\frac{q\mu}{2\pi}} \frac{7}{8}\frac{T^4}{6!}}\\
\int_0^\infty \frac{dE_p}{2\pi}  \prn{\frac{E_p}{2\pi}}^5 \brk{f_q + f_{-q}} &= {\frac{1}{6!}\prn{\frac{q\mu}{2\pi}}^6+\frac{1}{4!}\prn{\frac{q\mu}{2\pi}}^4 \frac{ T^2}{4!}+\frac{1}{2!}\prn{\frac{q\mu}{2\pi}}^2 \frac{7}{8}\frac{T^4}{6!}+\frac{31}{24}\frac{T^6}{8!}}\\
\end{split}
\end{equation}

Alternately, we can directly write down a generating function for these integrals by using the 
generating function of Bernoulli polynomials. Multiply eqn.\eqref{eq:gPoly} by $\tau^k$,
sum over $k=0$ to $\infty$ and then use eqn.\eqref{eq:BernoulliGen} to get 
\begin{equation}\label{eq:gGen}
\begin{split}
\int_0^\infty \frac{dE_p}{2\pi} \brk{g_q e^{\frac{\tau}{2\pi}E_p}+ g_{-q}e^{-\frac{\tau}{2\pi}E_p}} &= -\frac{2\pi}{\tau^2}
\brk{ \frac{\frac{\tau}{2}T}{\sin\prn{\frac{\tau}{2}T}}e^{\tau \frac{q\mu}{2\pi}} - \prn{1+\tau \frac{q\mu}{2\pi}}}\\
\end{split}
\end{equation} 
which we used in the main text to evaluate the integral in eqn.\eqref{eq:GEval}. In a similar vein
we get a generating function for the $f_q$ integral
\begin{equation}\label{eq:fGen}
\begin{split}
\int_0^\infty \frac{dE_p}{2\pi} \brk{f_q e^{\frac{\tau}{2\pi}E_p}-f_{-q}e^{-\frac{\tau}{2\pi}E_p}} &= \frac{1}{\tau}
\brk{ \frac{\frac{\tau}{2}T}{\sin\prn{\frac{\tau}{2}T}}e^{\tau \frac{q\mu}{2\pi}} - 1}\\
\end{split}
\end{equation} 

These formulae can be used to explicitly calculate $\bar{\mathcal{G}}_{anom}$. 
\begin{equation}
\begin{split}
\prn{\bar{\mathcal{G}}_{anom}}_{d=2} &= -2\pi \sum_{species}\chi_{_{d=2}} \brk{\frac{1}{2!}\prn{\frac{q\mu}{2\pi}}^2+\frac{T^2}{4!} }  u \\
\end{split}
\end{equation}
\begin{equation}
\begin{split}
\prn{\bar{\mathcal{G}}_{anom}}_{d=4} &=  -2\pi \sum_{species}\chi_{_{d=4}}\brk{\frac{1}{3!}\prn{\frac{q\mu}{2\pi}}^3+ 
 \prn{\frac{q\mu}{2\pi}}\frac{T^2}{4!}  } (2\omega)\wedge u \\
&\qquad-2\pi \sum_{species}\chi_{_{d=4}} \brk{\frac{1}{2!}\prn{\frac{q\mu}{2\pi}}^2+\frac{T^2}{4!} } (qB)\wedge u \\
\end{split}
\end{equation}
\begin{equation}
\begin{split}
\prn{\bar{\mathcal{G}}_{anom}}_{d=6} &=-2\pi \sum_{species} \chi_{_{d=6}} \brk{\frac{1}{4!}\prn{\frac{q\mu}{2\pi}}^4+\frac{1}{2!}\prn{\frac{q\mu}{2\pi}}^2 \frac{ T^2}{4!}+\frac{7}{8}\frac{T^4}{6!}} (2\omega)^2\wedge u\\
&\qquad-2\pi \sum_{species}\chi_{_{d=6}}\brk{\frac{1}{3!}\prn{\frac{q\mu}{2\pi}}^3+ 
 \prn{\frac{q\mu}{2\pi}}\frac{T^2}{4!}  } (2\omega)\wedge(qB)\wedge u \\
&\qquad-2\pi \sum_{species}\chi_{_{d=6}} \brk{\frac{1}{2!}\prn{\frac{q\mu}{2\pi}}^2+\frac{T^2}{4!} } \frac{(qB)^2}{2!}\wedge u \\
\end{split}
\end{equation}
\begin{equation}
\begin{split}
\prn{\bar{\mathcal{G}}_{anom}}_{d=8} &=  -2\pi \sum_{species} \chi_{_{d=8}} \brk{\frac{1}{5!}\prn{\frac{q\mu}{2\pi}}^5+\frac{1}{3!}\prn{\frac{q\mu}{2\pi}}^3 \frac{ T^2}{4!}+\prn{\frac{q\mu}{2\pi}} \frac{7}{8}\frac{T^4}{6!}}(2\omega)^3\wedge u\\
&\qquad -2\pi \sum_{species} \chi_{_{d=8}} \brk{\frac{1}{4!}\prn{\frac{q\mu}{2\pi}}^4+\frac{1}{2!}\prn{\frac{q\mu}{2\pi}}^2 \frac{ T^2}{4!}+\frac{7}{8}\frac{T^4}{6!}} (2\omega)^2\wedge(qB)\wedge u\\
&\qquad-2\pi \sum_{species}\chi_{_{d=8}}\brk{\frac{1}{3!}\prn{\frac{q\mu}{2\pi}}^3+ 
 \prn{\frac{q\mu}{2\pi}}\frac{T^2}{4!}  } (2\omega)\wedge\frac{(qB)^2}{2!}\wedge u \\
&\qquad-2\pi \sum_{species}\chi_{_{d=8}} \brk{\frac{1}{2!}\prn{\frac{q\mu}{2\pi}}^2+\frac{T^2}{4!} } \frac{(qB)^3}{3!}\wedge u \\
\end{split}
\end{equation}
\begin{equation}
\begin{split}
\prn{\bar{\mathcal{G}}_{anom}}_{d=10} &=-2\pi \sum_{species} \chi_{_{d=10}}\brk{\frac{1}{6!}\prn{\frac{q\mu}{2\pi}}^6+\frac{1}{4!}\prn{\frac{q\mu}{2\pi}}^4 \frac{ T^2}{4!}+\frac{1}{2!}\prn{\frac{q\mu}{2\pi}}^2 \frac{7}{8}\frac{T^4}{6!}+\frac{31}{24}\frac{T^6}{8!}} (2\omega)^4\wedge u\\
&\qquad -2\pi \sum_{species} \chi_{_{d=10}} \brk{\frac{1}{5!}\prn{\frac{q\mu}{2\pi}}^5+\frac{1}{3!}\prn{\frac{q\mu}{2\pi}}^3 \frac{ T^2}{4!}+\prn{\frac{q\mu}{2\pi}} \frac{7}{8}\frac{T^4}{6!}}(2\omega)^3\wedge(qB)\wedge u\\
&\qquad -2\pi \sum_{species} \chi_{_{d=10}} \brk{\frac{1}{4!}\prn{\frac{q\mu}{2\pi}}^4+\frac{1}{2!}\prn{\frac{q\mu}{2\pi}}^2 \frac{ T^2}{4!}+\frac{7}{8}\frac{T^4}{6!}} (2\omega)^2\wedge\frac{(qB)^2}{2!}\wedge u\\
&\qquad-2\pi \sum_{species}\chi_{_{d=10}}\brk{\frac{1}{3!}\prn{\frac{q\mu}{2\pi}}^3+ 
 \prn{\frac{q\mu}{2\pi}}\frac{T^2}{4!}  } (2\omega)\wedge\frac{(qB)^3}{3!}\wedge u \\
&\qquad-2\pi \sum_{species}\chi_{_{d=10}} \brk{\frac{1}{2!}\prn{\frac{q\mu}{2\pi}}^2+\frac{T^2}{4!} } \frac{(qB)^4}{4!}\wedge u \\
\end{split}
\end{equation}
The energy/charge/entropy currents can be obtained from these expressions via
\begin{equation}
\begin{split}
\bar{J}_{anom} &=-\frac{\partial \bar{\mathcal{G}}_{anom}}{\partial \mu}\\
\bar{J}_{S,anom}  &=-\frac{\partial \bar{\mathcal{G}}_{anom}}{\partial T} \\
\bar{q}_{anom} &= \bar{\mathcal{G}}_{anom}+T\bar{J}_{S,anom}+\mu \bar{J}_{anom}\\
\end{split}
\end{equation}

\section{Anomaly Polynomials}\label{app:anom}
One of the most succinct ways to capture the anomalies in a system is via the
anomaly polynomial $\mathcal{P}_{anom}(F,\mathfrak{R})$ of gauge field
strengths $F$ and the spacetime curvature $\mathfrak{R}$. We will not review how
these polynomials are calculated or how covariant anomalies are obtained from 
them since these topics are covered  well in various textbooks \cite{Weinberg:1996kr,Bertlmann:1996xk,Bastianelli:2006rx} and lecture notes \cite{Harvey:2005it,Bilal:2008qx}. We will mainly state the 
results relevant for Weyl fermions in arbitrary dimensions to save the reader
the effort of converting the standard results into our notation.

We will begin by reviewing various forms relevant for dealing with 
gravitational anomalies. Let $\mathfrak{R}_{ab}$ be the curvature 2-forms
of the spacetime with
\begin{equation}\label{eq:Rdef}
\mathfrak{R}_{ab} \equiv \frac{1}{2!}\ R_{abcd}\ dx^c\wedge dx^d 
\end{equation}

In $d=2n$ dimensions we can think of $\mathfrak{R}_{ab}$ as a real 
$2n\times 2n$ antisymmetric  matrix of 2-forms. At a give point in the
manifold, we can diagonalize it with the diagonal entries being  2-forms
\begin{equation}
 \mathfrak{R}_{ab}=\prn{\begin{array}{cccccc}
+i r_{_1} & 0 &\ldots& \ldots&\ldots& \ldots\\
0 &-i r_{_1} &\ldots & \ldots&\ldots& \ldots \\
\ldots& \ldots&\ldots& \ldots&\ldots& \ldots\\
\ldots& \ldots&\ldots& \ldots&\ldots& \ldots\\
\ldots& \ldots &\ldots & \ldots & +ir_n &0 \\
\ldots& \ldots &\ldots & \ldots & 0 & -ir_n \\
\end{array}}_{ab}
\end{equation}
where ${r}_{j=1,\ldots,n}$ are real 2-forms. Polynomials in these 2-forms can be used to
construct various other useful forms. The rest of this section basically consists
of various such polynomials, relations between them and their generating functions
etc. We start with the most basic form
\begin{equation}\label{eq:Rk}
\mathfrak{R}_k \equiv \frac{1}{2} \mathfrak{R}_{a_1}{}^{a_2}\wedge \mathfrak{R}_{a_2}{}^{a_3} \ldots  \wedge\mathfrak{R}_{a_k}{}^{a_1} =  \frac{1+(-1)^k}{2}\sum_j(ir_j)^k
\end{equation}
so $\mathfrak{R}_k$ is a is a $2k$-form which is a $k^{th}$-degree polynomial in the curvature 2-forms .
As is evident from the expression above, it is non-zero only when $k$ is even.

The next form which is we will introduce is called the $k^{th}$-Pontryagin class denoted by 
$p_{_k}(\mathfrak{R})$ which is $4k$-form using a specific $2k$-th degree polynomial in 
the curvature 2-forms. It is defined via the relation
\begin{equation}\label{eq:pk}
\begin{split}
\text{det}\brk{1+\frac{\tau}{2\pi}\mathfrak{R}} &= \prod_j\brk{1-\prn{\frac{\tau}{2\pi}ir_j}^2}  \equiv \sum_{k} \tau^{2k}p_{_k} (\mathfrak{R}) \\
\end{split}
\end{equation}
This gives
\begin{equation}
\begin{split}
p_{_1}(\mathfrak{R}) &= -\frac{1}{(2\pi)^2} \mathfrak{R}_2 \\
p_{_2}(\mathfrak{R}) &= -\frac{1}{(2\pi)^4} \brk{\frac{1}{2}\mathfrak{R}_4-\frac{1}{2}\mathfrak{R}_2^2} \\
p_{_3}(\mathfrak{R}) &= -\frac{1}{(2\pi)^6} \brk{\frac{1}{3}\mathfrak{R}_6-\frac{1}{2}\mathfrak{R}_2\mathfrak{R}_4+\frac{1}{6}\mathfrak{R}_2^3} \\
\end{split}
\end{equation}
which can be inverted to get 
\begin{equation}
\begin{split}
\frac{1}{(2\pi)^2} \mathfrak{R}_2 &= -p_{_1}(\mathfrak{R}) \\
\frac{1}{(2\pi)^4}\mathfrak{R}_4 &=-2p_{_2}(\mathfrak{R})+p^2_{_1}(\mathfrak{R}) \\
\frac{1}{(2\pi)^6}\mathfrak{R}_6 &=-3p_{_3}(\mathfrak{R}) + 3 p_{_1}(\mathfrak{R})p_{_2}(\mathfrak{R})-p^3_{_1}(\mathfrak{R})  \\
\end{split}
\end{equation}
We will express all the other polynomials in the basis of either $\mathfrak{R}_k$s or $p_{_k}(\mathfrak{R})$.

An object that plays an important role in the anomalies generated by Weyl fermions
is the  Dirac genus (or A-roof genus) $\hat{A}_k(\mathfrak{R})$  which is defined via
\begin{equation}
\begin{split}
\hat{A}\prn{\tau\mathfrak{R}} &\equiv \text{det}^{1/2}\brk{\frac{\frac{1}{2}\frac{\tau}{2\pi}\mathfrak{R}}{\sin\prn{\frac{1}{2} \frac{\tau}{2\pi}\mathfrak{R}}}} = \prod_j\brk{\frac{\frac{1}{2}\frac{\tau}{2\pi}r_j}{\sinh\prn{\frac{1}{2} \frac{\tau}{2\pi}r_j}}}= \sum_k \tau^{2k} \hat{A}_{_k}(\mathfrak{R}) \\
\end{split}
\end{equation}
which gives
\begin{equation}\label{eq:AExp} 
\begin{split}
\hat{A}_{_1}(\mathfrak{R})&=\frac{1}{(2\pi)^2} \brk{\frac{\mathfrak{R}_2}{4!}} = \frac{1}{4!}\brk{- p_{_1}(\mathfrak{R})} \\
\hat{A}_{_2}(\mathfrak{R})&=\frac{1}{(2\pi)^4} \brk{\frac{1}{4}\prn{\frac{\mathfrak{R}_4}{6!}}+\frac{1}{2}\prn{\frac{\mathfrak{R}_2}{4!}}^2}
=\frac{1}{6!(2\pi)^4}\frac{1}{8}\prn{2\mathfrak{R}_4+5\mathfrak{R}_2^2}\\
&=\frac{1}{6!} \brk{-\frac{1}{2}p_{_2}(\mathfrak{R})+\frac{7}{8}p^2_{_1}(\mathfrak{R})}\\
\hat{A}_{_3}(\mathfrak{R})&=\frac{1}{(2\pi)^6}\brk{
\frac{2}{9}\prn{\frac{\mathfrak{R}_6}{8!}}+ \frac{1}{4}\prn{\frac{\mathfrak{R}_2}{4!}}\prn{\frac{\mathfrak{R}_4}{6!}}
+\frac{1}{6}\prn{\frac{\mathfrak{R}_2}{4!}}^3 }  \\
&= \frac{1}{8!(2\pi)^6}\frac{1}{24}\prn{
\frac{16}{3}\mathfrak{R}_6+14\ \mathfrak{R}_4\mathfrak{R}_2+\frac{35}{3}\mathfrak{R}_2^3
}\\
&= \frac{1}{8!}\brk{
-\frac{2}{3}p_{_3}(\mathfrak{R}) +\frac{11}{6}p_{_1}(\mathfrak{R})p_{_2}(\mathfrak{R})
 -\frac{31}{24}p^3_{_1}(\mathfrak{R}) } \\
\end{split}
\end{equation}
The $k^{th}$-Dirac genus  $\hat{A}_{_k}(\mathfrak{R})$ is hence a $4k$-form using a specific $2k$-th degree polynomial in 
the curvature 2-forms.

We now want to calculate the Dirac genus in the special case where the only non-zero Pontryagin
class is $p_{_1}(\mathfrak{R})$ with $p_{_{k>1}}(\mathfrak{R})=0$. It is clear that in this case
$\hat{A}_{_k}(\mathfrak{R})\sim p_{_1}^k(\mathfrak{R})$ but we want to calculate the numerical
coefficient exactly. To do this, we use the following trick : we set the 2-forms $r_{j>1}=0 $
with a non-zero $r_1$.
Then
\[ \text{det}\brk{1+\frac{\tau}{2\pi}\mathfrak{R}} =1-\prn{\frac{\tau}{2\pi}ir_1}^2  \]
which means 
\[p_{_1}(\mathfrak{R}) =  \prn{\frac{r_1}{2\pi}}^2  \ ,\quad p_{_{k>1}}(\mathfrak{R}) =  0  \]
The roof-genus is easily evaluated in this case as 
\[ \frac{\frac{1}{2}\frac{\tau}{2\pi}r_j}{\sinh\prn{\frac{1}{2} \frac{\tau}{2\pi}r_j}}= \frac{\frac{\tau}{2}\sqrt{p_{_1}(\mathfrak{R})}}{\sinh\prn{\frac{\tau}{2}\sqrt{p_{_1}(\mathfrak{R})}}} = \frac{\frac{\tau}{2}\sqrt{-p_{_1}(\mathfrak{R})}}{\sin\prn{\frac{\tau}{2}\sqrt{-p_{_1}(\mathfrak{R})}}}\]
So this gives us the required answer 
\begin{equation}
\brk{\hat{A}\prn{\tau\mathfrak{R}}}_{ p_{_{k>1}}(\mathfrak{R}) =  0} = \frac{\frac{\tau}{2}\sqrt{-p_{_1}(\mathfrak{R})}}{\sin\prn{\frac{\tau}{2}\sqrt{-p_{_1}(\mathfrak{R})}}}
\end{equation}

The anomaly polynomial for a bunch of Weyl fermions in a $d=(2n-1)+1$ 
dimensional spacetime is given by 
\[ \prn{\mathcal{P}_{anom}}_{d=2n}=- 2\pi\sum_{species}  \chi_{_{d=2n}}\ \brk{\hat{A}\prn{\tau \mathfrak{R}}\ 
e^{\frac{\tau}{2\pi}qF}}_{\tau^{n+1}}\]
where we have assumed  that there is a single $U(1)$ symmetry under which the charge of the
Weyl fermion is denoted by the letter $q$. The sum is over the species which means that 
each particle/anti-particle pair contributes one term to the sum (the answer does not depend on
whether we take the chirality/charge of a particle or the anti-particle ). We now proceed to 
write down the explicit expressions for $\mathcal{P}_{anom}$ by using the formula above.
\begin{equation}\label{eq:PExp2d}
\begin{split}
(\mathcal{P}_{anom})_{_{d=1+1}} = -2\pi \sum_{species}  \chi_{_{d=2}}&\brk{\frac{1}{2!}\prn{\frac{qF}{2\pi}}^2+\hat{A}_1(\mathfrak{R}) }\\
(\mathcal{P}_{anom})_{_{d=3+1}} = -2\pi\sum_{species}  \chi_{_{d=4}} &\brk{\frac{1}{3!}\prn{\frac{qF}{2\pi}}^3+ 
 \prn{\frac{qF}{2\pi}}\hat{A}_1(\mathfrak{R}) } \\
(\mathcal{P}_{anom})_{_{d=5+1}}= -2\pi\sum_{species}  \chi_{_{d=6}} & \left[\frac{1}{4!}\prn{\frac{qF}{2\pi}}^4+\frac{1}{2!}\prn{\frac{qF}{2\pi}}^2 \hat{A}_1(\mathfrak{R}) 
+\hat{A}_2(\mathfrak{R})\right]\\
(\mathcal{P}_{anom})_{_{d=7+1}} = -2\pi\sum_{species}  \chi_{_{d=8}} & \left[\frac{1}{5!}\prn{\frac{qF}{2\pi}}^5+\frac{1}{3!}\prn{\frac{qF}{2\pi}}^3 \hat{A}_1(\mathfrak{R})
+\prn{\frac{qF}{2\pi}} \hat{A}_2(\mathfrak{R}) \right]\\
(\mathcal{P}_{anom})_{_{d=9+1}} = -2\pi\sum_{species}  \chi_{_{d=10}} & \left[\frac{1}{6!}\prn{\frac{qF}{2\pi}}^6+\frac{1}{4!}\prn{\frac{qF}{2\pi}}^4\hat{A}_1(\mathfrak{R})\right.\\
&\qquad\left.+\frac{1}{2!}\prn{\frac{qF}{2\pi}}^2\hat{A}_2(\mathfrak{R}) +\hat{A}_3(\mathfrak{R})\right]\\
\end{split}
\end{equation}
where explicit expressions for $\hat{A}_i(\mathfrak{R})$ are given by eqn.\eqref{eq:AExp} 

\section{Chiral magnetic effect in arbitrary dimensions}\label{app:magnetic}
In this appendix, our main aim is to understand in more physical terms how the 
boundary condition for the chiral spectral current arises. In particular, we 
would like to understand the rate at which the zero modes are injected into the
fluid in terms of Landau level physics.

Let us consider a Weyl fermion in $(2n-1)+1$ dimensions. As before we have
a finite temperature $T$, a chemical potential $\mu$ and a constant velocity
$u^\mu$. We will again work in the rest frame of $u^\mu$. The main aim of this
section is to understand the transport that arises when you turn on a magnetic 
field in this rest frame. To do this, let us divide the $2n-1$ spatial directions
into a direction $x_1$ and $n-1$ planes  where the $k^{th}$ plane is the
$(x_{i_{2k}},x_{i_{2k+1}})$ plane where $k=1,\ldots,n-1$. We 
by consider the system with a uniform  magnetic field strength turned on in all these
$n-1$ spatial planes $b_k\equiv B_{i_{2k} i_{2k+1}}\neq 0$ for $k=1,\ldots,n-1$ 
leaving out one spatial direction $x_1$. 

We can solve the Weyl equation in this magnetic background and the solutions are the 
well-known Landau levels with their Landau degeneracies. The energy levels are determined
by the charge $q$, the momentum along the $x_1$ direction $p_1$ and 
the spin along $k^{th}$ plane $S_{i_{2k} i_{2k+1}}\equiv\frac{1}{2}\sigma_k$  (the
allowed values are $\sigma_k=\pm 1$ ) and the Landau level number in the $k^{th}$
plane being a non-negative integer $n_k$. 
\begin{equation}
\begin{split}
E^2 &\equiv p_1^2 + 2\sum_{k=1}^{n-1}\brk{ |qb_k|\prn{n_k+\frac{1}{2}}-\frac{1}{2}q b_k \sigma_k }\\
\end{split}
\end{equation}
where we see the free-particle dispersion along $x_1$ direction, the harmonic oscillator like
spectrum of the covariant Laplacian along with the Zeeman splitting due to the magnetic moment. 
The Landau degeneracy in momentum between $p_1$ and $p_1+dp_1$ (for a fixed  
Landau level number $n_k$ and spins $\sigma_k$) is 
\begin{equation}
\begin{split}
\mathcal{D}_{\text{Landau}} = \frac{dp_1}{2\pi}\prod_{k=1}^{n-1} \frac{|qb_k|}{2\pi}
\end{split}
\end{equation}
We notice that these are effectively a collection of $1+1$ dimensional free Dirac/Weyl Fermions 
with different masses. So one can repeat the analysis in section \S\ref{sec:2d} and we conclude 
that there should be a current in this rest frame (along $x_1$ direction)
which just depends on the chiral states.

To get a chiral state one has to set the effective $2d$ mass appearing above to zero.
: it is easy to see that the 2d-Weyl states are obtained only when 
$n_k=0,\sigma_k = \text{sign}\prn{q b_k}$. These states have $E^2=p_1^2 \equiv E_p^2$
and a degeneracy 
\[  \frac{dp_1}{2\pi}\prod_{k=1}^{n-1} \frac{|qb_k|}{2\pi}=\frac{dE_p}{2\pi}\prn{\prod_{k=1}^{n-1} \frac{qb_k}{2\pi}} \prn{\prod_{k=1}^{n-1} \sigma_k }\]

If we denote the $1+1$-dimensional chirality as $\chi_{_{d=2}}$ their total contribution to $\mathcal{G}^\mu_{anom}$ is
\begin{equation}
\begin{split}
&(e_1)^\mu\prn{\prod_{k=1}^{n-1} \frac{qb_k}{2\pi}} \chi_{_{d=2}}\prn{\prod_{k=1}^{n-1} \sigma_k } \times\int_0^\infty \frac{dE_p}{2\pi} g_q\\
\end{split}
\end{equation}
where $(e_1)^\mu$ is the unit vector along $x_1$ direction which ensures that 
the current is along the $x_1$ direction. We recognize the appearance of the 
$d=2n$ chirality
\[ \chi_{_{d=2n}} = \chi_{_{d=2}}\prod_{k=1}^{n-1} \sigma_k  = \chi_{_{d=2}}\prod_{k=1}^{n-1} \text{sign}\prn{q b_k}  \]
and this equation fixes the 2d chirality $\chi_{_{d=2}}$ of the zero Landau-level  in terms
of the chirality $\chi_{_{d=2n}}$ of the original Weyl fermion. This means we get one 2d-Weyl
fermion of definite chirality for each Weyl fermion in higher dimensions.

As before, it is useful to take a Hodge dual and write this in terms of forms. Let $F_{\mu\nu}$ be
the field strength which we can think of as a 2-form $F$. The rest frame electric field is 
$E_\mu = u^\nu F_{\mu\nu}$  which we can think of as a 1-form $E$. The rest frame magnetic field
is a 2-form obtained by subtracting the electric part from $F$, 
\[ B\equiv F-u\wedge E \]
where $u=u_\mu dx^\mu$ is the velocity 1-form. In terms of these quantities, we finally get the
total contribution from all fermions as
\begin{equation}
\begin{split}
(\bar{\mathcal{G}}_{anom})_{u=\text{constant}} &= \sum_F\int_0^\infty \frac{dE_p}{2\pi} g_q\ \chi_{_{d=2n}}\prn{ \frac{qB}{2\pi}}^{n-1}\wedge \frac{u}{(n-1)!}\\
\end{split}
\end{equation}
where the sum is performed over all the fermionic particles
counting the particle and the anti-particle separately. We have denoted in the subscript
our assumption that the velocity $u^\mu$ is constant. This gives 
the rate of injection of chiral zero modes by the magnetic field as 
\begin{equation}
\begin{split}
\bar{\mathcal{J}}_q(E_p=0) &= 1 \frac{\chi_{_{d=2n}}}{2\pi}    \prn{\frac{qB}{2\pi}}^{n-1}\wedge \frac{u}{(n-1)!}\\
\end{split}
\end{equation}

Certain comments are in order - when we turn on a mild magnetic field over a fluid, the
gap between the Landau levels (and the Zeeman split) are small and hence one expects that
any transport will get contribution from many Landau levels not just the lowest Landau level.
This expectation is in fact correct in the case of conventional quantities like
pressure or energy density - but remarkably enough the anomaly-induced contribution 
above does not get any contribution from the higher Landau levels even when the 
magnetic field is small. This is related to the statement that only 
chiral fermions contribute to the sum above . As a consistency check, 
one can  easily check that Dirac fermions do not contribute 
to this sum ( this follows  since in a Dirac species for a given
value of $q$ , $\chi_{_{d=2n}}$ takes both 
the values $\pm 1$ thus canceling out in the above sum ).

 \section{Notation}\label{app:notation}
We work in the $(-++\ldots)$ signature. The dimensions of the spacetime in which the fluid lives is denoted by $d=2n$.
The Greek indices $\mu,\nu= 0,1,\ldots,d-1$ are used as space-time indices.

We denote Hodge-duals by an overbar - for example, $\bar{J}$ is the 2n-1 form Hodge-dual to the
1-form $J_\mu$. We mostly just use the Hodge-duality between 1-forms and 2n-1 forms and our conventions
are completely defined by  the following statement- given any $2n-1$ form $\bar{V}$ hodge-dual to $V_\mu$
and a 1-form $A_\mu$, we have 
\begin{equation}
\begin{split}
D\bar{V}&=(D_\mu V^\mu)\ \text{Vol}_{2n}\\
A\wedge\bar{V}&=-\bar{V}\wedge A = A_\mu V^\mu \ \text{Vol}_{2n}\\
\end{split}
\end{equation}
Given a 0-form $\alpha$ its Hodge-dual 2n-form is simply $\bar{\alpha}\equiv \alpha \ \text{Vol}_{2n}$.  

We have included a table with other useful parameters used in the text. In the table~\ref{notation:tab}, the relevant equations are denoted by their respective equation numbers appearing inside parentheses.

\newpage
\begin{table}\label{notation:tab}
 \centering
 \begin{tabular}{||r|l||r|l||}
   \hline
   \multicolumn{4}{||c||}{\textbf{Table of Notation}} \\
   \hline 
   Symbol & Definition & Symbol & Definition \\
   \hline
   $\varepsilon$ & Energy density & $p$ & Pressure \\
   $n$ & Charge density & $s$ & Entropy density \\
   $\mu$& Chemical potential  & $T$ & Temperature\\
   $\beta$ & $1/T$ & $\chi_{_{d=2n}}$ & 2n dimensional chirality \\
   $q$ & Fermion charge &  $E_p$ & Fermion energy \\  
  $g_q$ & $ -\frac{1}{\beta}\ln\brk{1+e^{-\beta(E_p-q\mu)}}$  & $f_q$ & $\brk{e^{\beta(E_p-q\mu)}+1}^{-1}$\\
  $\mathfrak{R_{ab}}$  & Curvature 2-forms\eqref{eq:Rdef} & $\mathfrak{R}_k$ & See \eqref{eq:Rk}\\ 
  $\mathcal{P}_{anom}$ & Anomaly polynomial & $p_{_k}(\mathfrak{R})$ & Pontryagin class \eqref{eq:pk}\\
  $\mathfrak{F}^\omega_{anom}$ & See \eqref{eqn:Fw} & $\hat{A}_{_k}(\mathfrak{R})$ & A-roof genus \eqref{eq:AExp} \\
   \hline
   $u^\mu,u$ & Fluid velocity,   &  $a_\mu,a$ & Acceleration field \\
             & 1-form            &            &  $(u.D)u_\mu$, 1-form \\
   $g_{\mu\nu}$ & Spacetime metric & $P_{\mu\nu}$ & $g_{\mu\nu}+u_\mu u_\nu$ \\
  $\sigma_{\mu\nu}$ & Shear strain rate  & $\omega_{\mu\nu},\omega$ & Fluid vorticity, 2-form \\

   \hline
   $T^{\mu\nu}$ & Energy-momentum  & $J^{\mu}$ & Charge currents\\
    & tensor of the fluid &  &  with anomalies \\
   $J^\mu_S $ & Entropy current &  $D$ & Exterior derivative\\
   $\mathcal{G}^{\mu}_{anom}$ & Anomaly-induced  & $\bar{\mathcal{G}}_{anom}$ & Hodge-dual of $\mathcal{G}^{\mu}_{anom}$  \\
                    & Gibbs current    &  & $2n-1$ form \\
   $q^{\mu}_{anom}$ & Anomaly-induced  & $\bar{q}_{anom}$ & Hodge-dual of $q^{\mu}_{anom}$  \\
                    & heat current    &  & $2n-1$ form \\
    $J^{\mu}_{anom}$ & Anomaly-induced & $\bar{J}_{anom}$ & Hodge-dual of $J^{i\mu}_{anom}$ \\
        & Charge current &  & $2n-1$ form \\
   $J^\mu_{S,anom} $ & Anomaly-induced  & $\bar{J}_{S,anom}$ &  Hodge-dual of $J^\mu_{S,anom} $\\
                     &Entropy current &  & $2n-1$ form \\
   $F_{\mu\nu},F$ & non-dynamical gauge   & $E^\mu,E$ & Rest frame electric \\
               & field strength, 2-form &  & field $F_{\mu\nu}u^\nu$, 1-form \\
   $B_{\mu\nu},B$ & Rest frame magnetic & $\mathfrak{A}$ & Anomaly   \\
               &  fields $F-u\wedge E$ &  &  $D\bar{J}\equiv\bar{\mathfrak{A}}$ \\
  $\mathcal{J}^\mu_q$ & Chiral Spectral current & $\bar{\mathcal{J}}_q$ & Hodge-dual of $\mathcal{J}^\mu_q$\\
  $\mathcal{J}^E_q$ & See eqn\eqref{eq:JE} & $\bar{\mathcal{J}}^E_q$ & Hodge-dual of $\mathcal{J}^E_q$\\
  \hline
\end{tabular}
\end{table}

\bibliographystyle{JHEP}
\bibliography{chiralZ}
\end{document}